\def\,{\thinspace}

\documentclass[twocolumn]{aastex7}
\usepackage{CJKutf8}
\usepackage{amsmath}
\usepackage{booktabs}
\usepackage{graphicx}  
\usepackage{adjustbox}
\usepackage{tabularx}
\usepackage{caption}
\usepackage{longtable}
\usepackage{needspace}
\usepackage{multirow}

\begin{document}
\begin{CJK}{UTF8}{bsmi}
\title{Disclosing Submillimeter Galaxy Formation: Mergers or Secular Evolution?}

\author[orcid=0009-0009-5330-3287, sname="Chan"]{Siu-Wang Chan({陳瀟弘})}
\affiliation{Purple Mountain Observatory, Chinese Academy of Science, 10 Yuanhua Road, Nanjing 210023, People’s Republic of China}z
\affiliation{School of Astronomy and Space Sciences, University of Science and Technology of China, Hefei 230026, People’s Republic of China}
\email[show]{xhchen@pmo.ac.cn}

\author[orcid=0000-0003-3139-2724, sname="Ao"]{Yiping Ao}
\affiliation{Purple Mountain Observatory, Chinese Academy of Science, 10 Yuanhua Road, Nanjing 210023, People’s Republic of China}
\affiliation{School of Astronomy and Space Sciences, University of Science and Technology of China, Hefei 230026, People’s Republic of China}
\email[show]{ypao@pmo.ac.cn}

\author[orcid=0000-0003-3032-0948, sname='Tan']{Qinghua Tan}
\affiliation{Purple Mountain Observatory, Chinese Academy of Science, 10 Yuanhua Road, Nanjing 210023, People’s Republic of China}
\affiliation{School of Astronomy and Space Sciences, University of Science and Technology of China, Hefei 230026, People’s Republic of China}
\email[]{}

\begin{abstract}
    We analyze the morphology of 125 submillimeter galaxies (SMGs) in the PRIMER-COSMOS field using double Sérsic modeling on JWST NIRCam images across six bands (F150W, F200W, F277W, F356W, F410M and F444W), with SMGs being classified by bulge Sérsic index ($n\_bulge$) and bulge-to-total luminosity ratio (B/T). The Kolmogorov–Smirnov test between the bright ($SFR>175~M_\odot ~yr^{-1}$) and the faint group ($SFR<175~M_\odot~yr^{-1}$) reveals no significant statistical differences in morphology across bands. However, we notice that SMGs skew towards higher B/T ratios and lower $n\_bulge$ from shorter to longer wavelengths. In F444W, bright SMGs exhibit higher B/T and lower $n\_bulge$, indicating flatter , disturbed bulges, while faint SMGs show lower B/T and higher $n\_bulge$.  Notably, SMGs with higher $B/T $ tend to have low Sérsic, challenging the local universe dichotomy of classical bulges ($B/T > 0.5$, $n > 4$) versus pseudo-bulges ($B/T < 0.35$, $n < 2$). In the F277W , non-parametric morphological measurements indicate predominantly disk-dominated patterns, with only $24\%$ of SMGs demonstrating merger signatures. After the removal of SMGs with disturbed morphology , the bulge classification scheme in F277W shows pseudo-bulges($21\%$) and  clump migration bulges($16\%$) from secular evolution , compared to $4\%$ merger-built bulges. Surprisingly , $48\%$ of SMGs defy the classification scheme , showing high $B/T$($\sim0.7$) but low Sérsic index($n\_bulge \leq 1$). Bars are confirmed in $7 \%$ of SMGs. This work suggests that secular evolution takes precedence over major mergers , supporting the idea that isolated evolution fueled by filamentary gas inflow plays a non-negligible role in the formation of SMGs. 
\end{abstract}



\section{Introduction}
It is well known in the field that sub-millimeter galaxies play one of the most important roles at "cosmic noon"($z\sim 2-3$) , due to the fact that these systems harbor half of the star formation activities occurring over cosmic time , which are enshrouded by heavy dust\citep{1996A&A...308L...5P, 1998ApJ...508...25H, 2020MNRAS.494.3828D}. The dusty systems with high star formation rate ($SFR\sim10^{2-3}M_\odot ~ yr^{-1}$)systems ,  with extreme luminosity at far-IR and sub-mm wavelength($L_{IR}\sim10^{12-13}L_\odot$) , containing a gas reservoirs of $M_{gas}\sim 10^{10-11}M_\odot$ \citep{2002PhR...369..111B, 2006ApJ...640..228T, 2010ApJ...720L.131R} , are first uncovered by the Submillimeter Common-User Bolometer Array(SCUBA) at 850 $\mu m$ on the James Clark Maxwell Telescope(JCMT) in the late 1990s\citep{1997ApJ...490L...5S, 1998Natur.394..241H, 1998Natur.394..248B} , revolutionizing the way of studying extragalactic astronomy.

Numerous follow-up studies have been carried out across multiple bands , spanning optical to radio , to paint a full picture of these mysterious high-z galaxies throughout the decades. The heavy dust permeating among the plane of galactic disks causes the invisibility of SMGs at rest-frame UV/optical band , in the operation by either the space- and/or ground-based facilities , complicating the morphological confirmation\citep{2015ApJ...806..110D} , which results in several different naming variations for this type of galaxies, for example the "Dusty star forming galaixes(DSFGs)" , "HST-faint" , and "NIR-Dark"\citep{2004ApJ...614..671C, 2014ApJ...788..125S, 2021MNRAS.502.3426S} etc.

Due to its intense star formation activities and high infrared luminosity , it is natural for people to associate SMGs to ultraluminous infrared galaxies(ULIRGs) in the local universe\citep{1996ARA&A..34..749S, 2006asup.book..285L}, suspecting similar formation patterns for these high-z systems. The majority of the ULIRGs($\sim 90\%$) are mostly assembled by major mergers , in which extreme starbursts are caused by gas compression , while for the SMGs , although previous studies using the Advanced Camera for Surveys (ACS) and Wide Field Camera 3 (WFC3) on the Hubble Space Telescope(HST) indicate $50-60\%$ of merger rate\citep{2012ApJ...757...23K, 2015ApJ...799..194C, 2003ApJ...596L...5C, 2004ApJ...616...71S} , recent high-resolution images from the JWST show only $\sim20\%$ of the SMGs exhibiting tidal features\citep{2024A&A...691A.299G, 2023A&A...676A..26G} , perplexing previous oversimplifications of their evolutionary pathways.

One of the most characteristic and complicated part of the SMGs is the dust-surrounded starburst core lied in the center of the galaxy , detected by the James Clark Maxwell Telescope(JCMT), the South Pole Telescope(SPT) , and the Atacama Large Millimeter/submillimeter Array(ALMA)\citep{2016ApJ...833..103H, 2019MNRAS.490.4956G} , which plays an essential part in disclosing the formation truth and has yet to be fully understood. Several conjunctures have been brought up to account for the dilemma of formation mechanism: 1) major mergers driven , 2) violent disk instabilities , causing the clumps to migrate to the center , 3) cold gas accretion from the galaxy filaments(secular evolution). The major-merger origin (galaxy to galaxy mass ratio  $\geq1:4$) , is supported by the HST merger rate while being opposed by the merger rate derived based on the JWST data , possibly due to the interference of heavy dust extinction in previous studies\citep{2012ApJ...757...23K, 2015ApJ...799..194C, 2010MNRAS.405..234S}. Whereas the minor mergers , can indeed sustain graduate star formation, but yet being questioned for limited dynamical impact\citep{2019ApJ...876..130H, 2015Natur.525..496N}. As for gas accretion , it serves as a good explanation for the disk-dominated morphology at high-z ,  but still struggles to explain the ultra-luminous SMGs\citep{2013MNRAS.428.2529H}. None of these three monopolies on evolution mechanisms can fully account for the 
reality, suggesting a co-existence varying with luminosities and environments\citep{2005MNRAS.356.1191B, 2010MNRAS.404.1355D, 2014PhR...541...45C}.

In this work , we analyze the morphological properties of 125 statistically robust ALMA-detected SMGs in \texttt{PRIMER-COSMOS} field from Public Release IMaging for Extragalactic Research(PRIMER, Program ID:1837, PI: James S. Dunlop) , via two-dimensional decomposition and non-parametric analysis, as long as bar identification. The observational data are all presented here as part of the DAWN JWST Archive(DJA)\citep{2023ApJ...947...20V} , which includes fully reduced and pre-processed data using $grizli$ pipeline\citep{2023zndo...8370018B} for all HST and JWST imaging data.

The structure of this paper is outlined below. Section 2 details the galaxy sample along with the photometric data. In Section 3 , we explain our methods for determining the galaxy morphology , including two-dimensional decomposition , merger identification , and bar identification strategies. Section 4 illustrates the results of our analysis , and further discussions are provided in Section 5. Throughout this paper , we assume that the Planck cosmology:$H_0 = 70.0 \ km.s^ {-1}$ , $\Omega_M=0.3$ , $\Omega_\lambda=0.7$.


\section{Data \& Sample selection}
To construct a sample of ALMA-detected galaxies with robust multi-wavelength coverage , we utilize the $A^3COSMOS$ \citep{2019ApJS..244...40L, 2024A&A...685A...1A} robust galaxy catalog\footnote{\url{https://uni-bonn.sciebo.de/s/0LrfuA14s2YP3cz}} as our parent catalog. The $A^3COSMOS$ project leverages archival ALMA (Atacama Large Millimeter/submillimeter Array) observations to study galaxy evolution , and the included catalog contains galaxies with secure ALMA detections (peak $S/N > 5.4$ , lowest $S/N> 3$), with optical/near-infrared counterparts , and derived physical properties like stellar mass , redshift , and star formation rate(SFRs) etc. The $A^3COSMOS$ catalog incorporates 1.25 mm observations from 173 individual ALMA projects across 3233 pointings in the COSMOS field. 
\begin{figure}[htbp]
    \centering
    \includegraphics[width=1.01\linewidth]{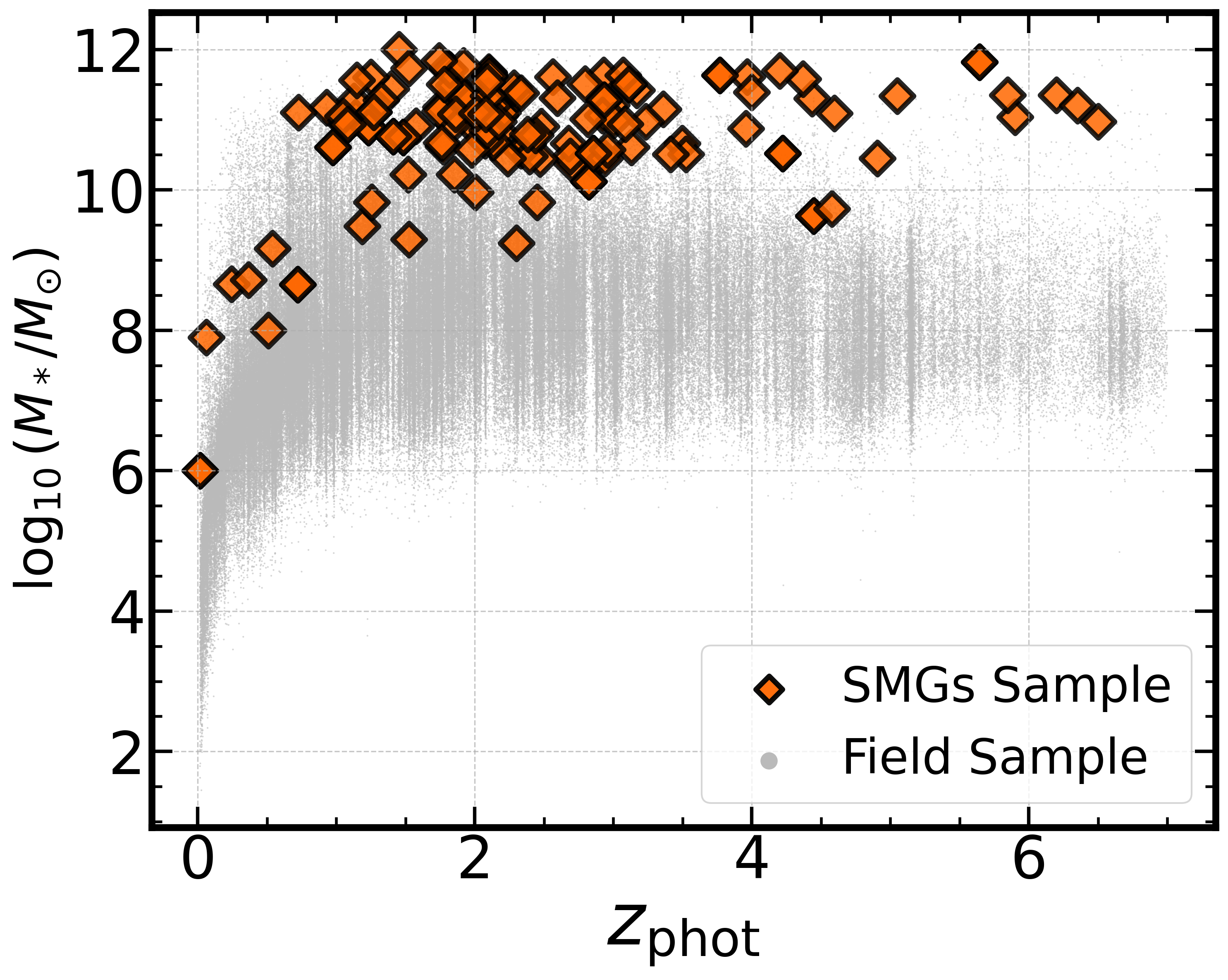}
    \caption{Photometric Redshift vs Stellar Mass for 125 selected SMGs(orange) , and the field sample(gray) in PRIMER-COSMOS field}
    \label{fig:smgs_vs_field}
\end{figure}

We then cross-match the ALMA positions for the SMGs with the photometric catalogs provided by the Dawn JWST Archive(DJA)\citep{2023ApJ...947...20V} of The Public Release IMaging for Extragalactic Research (PRIMER; Dunlop et al. 2021) survey.  As a multi-wavelength , multi-instrument survey of UDS and COSMOS covering 234 and 144 arcmin$^2$ respectively , this extensive dataset provides deep NIRCam imaging. The observation consists of the seven wide-band and one medium-band NIRCam filter(F090W, F115Wt, F150W, F200W, F277W, F356W, F410M, F444W) , along with two wide-band MIRI filters(F770W, F1800W). Additionally, a portion of the PRIMER-COSMOS field is covered by HST ACS and WFC3 coverage from CANDELs\citep{2011ApJS..197...35G, 2011ApJS..197...36K}, providing optical-to-near-infrared data ranging from 0.4$\mu m$ to 1.6$\mu m$. We excluded filters suffering from the issues of invisibility of the near-infrared counterparts (F090W, F115W) , and those with insufficient resolution for further analysis(F770W, F1800W) , which results in the remaining six bands(F150W, F200W, F277W, F356W, F410M, F444W) with robust detections of the JWST. Since detections in differenct wavelength, we convert the flux detected in different bands to $870\mu m$ via $f_{870 \mu m}/f_{obs}=(\nu_{870\mu m}/\nu_{obs})^3$. 

In the end , we obtained 125 unique sources detected by ALMA, with the complete catalog 
 \ref{tab:alma_jwst} in the Appendix. The median of $log_{10}(M_*/M_\odot)$ is $ 10.98_{-0.58}^{+0.53} $ (16th to 84th percentile) , and the median redshift for our sample is $ 2.29_{-1.48}^{+3.89} $(16th to 84th percentile) , which is consistent with the typcial redshift range of the SMGs but skews toward higher redshift, dominating fainter end of the SMGs population($S_{870 \mu m}=1.88_{-0.72}^{+1.28} ~ mJy$). The physical properties presented in this paper are all retrieved from the $A^3COSMOS$ project \citep{2024A&A...685A...1A}, in which they performed SED fitting with {MAGPHYS}, integrating ALMA sub-mm data with extensive ancillary data , including radio (e.g., VLA 1.4 GHz and 3 GHz), far-infrared (e.g., Herschel, Spitzer/MIPS 24 µm), and optical-to-near-infrared photometry.  The uncertainties of the photometric redshift conform to $0.06(1+z)$ , which is suitable for statistical measurements but not clusters proximity survey. As for stellar mass , the A3COSMOS paper excludes outliers and unreliable associations , ensuring a robust catalog with a minimized chi-square error. The distribution of the redshift versus the stellar mass is shown in Figure \ref{fig:smgs_vs_field}. The median SFR is $185 ~ M_\odot ~ yr^{-1}$ , with a 16th to 84th percentile ranging from $55-795 ~ M_\odot ~ yr^{-1}$. 

\section{Methodologies} \label{sec:floats}
In this section , we exhibit our methods to detail the morphological information of SMGs , including initial data reduction , two-dimensional(2D) parametric decomposition modeling , non-parametic approaches , along with bar identification techniques.

\begin{figure*} 
    \centering
    \includegraphics[width=1\linewidth]{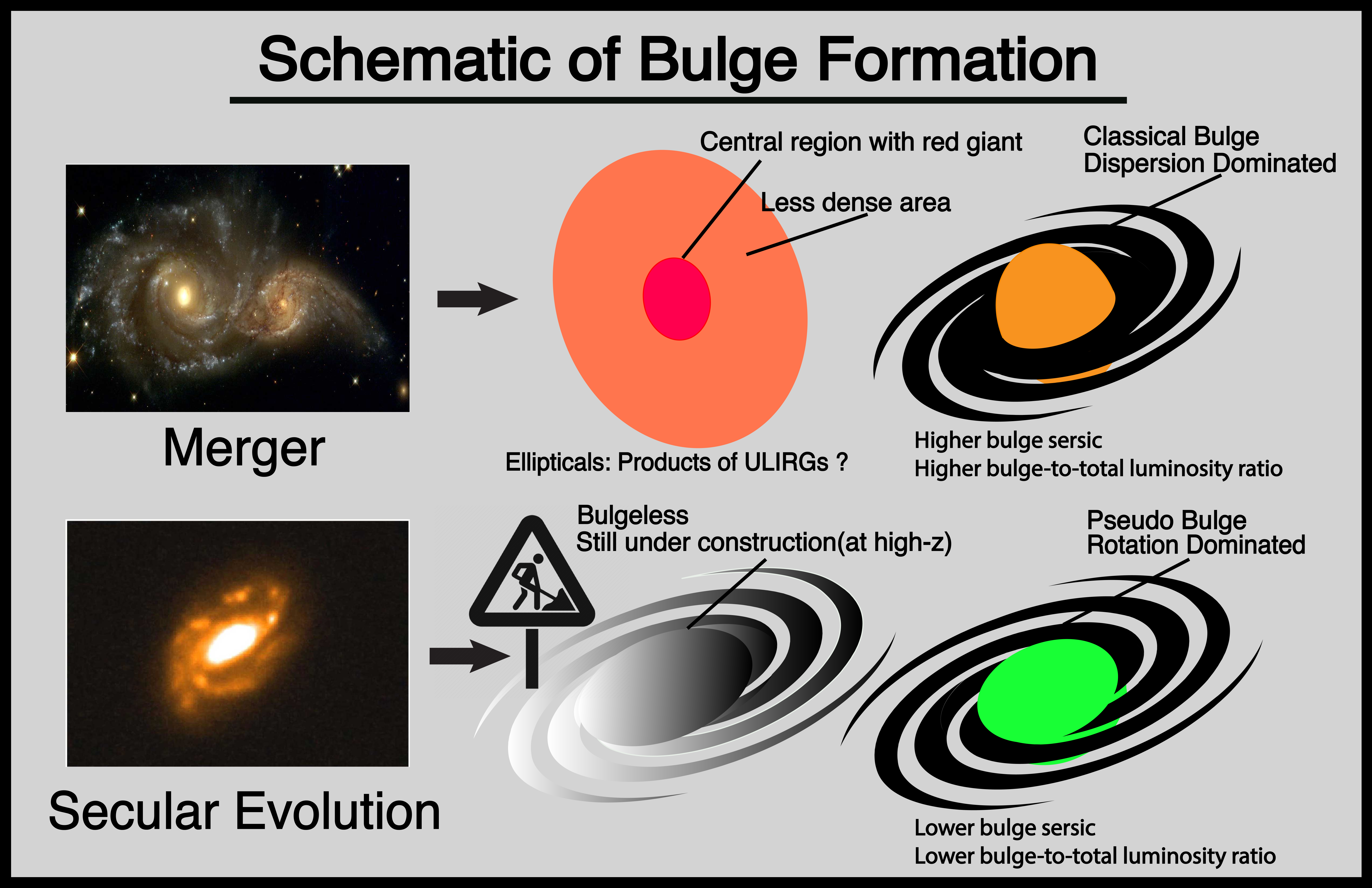}
    \caption{Simple assumptive sketch based on previous studies summarizing the bulge formations in submillimeter galaxies presented in this paper , in which the classical bulge and ellipticals(kinematically hot, high $B/T$, high Sérsic) are the products of mergers whereas the pseodobulges and the bulgeless galaxies(kinematically cold, low $B/T$, low Sérsic) are fabricated by degrees in the secular evolution.}
    \label{fig:merger_secular}
\end{figure*}

\subsection{Data reduction}

Firstly , we use the preprocessed $JWST$ data provided by the Dawn JWST Archive (DJA). The data reduction pipeline , as detailed in \citep{2023ApJ...947...20V}, processes level-2 NIRCam products with GRIZLI , applying zero-point corrections and artifact removal, then aligns them with the HST Complete Hubble Archive for Galaxy Evolution \citep{2022ApJS..263...38K} data to Gaia DR3\citep{2021A&A...649A...1G} at $0''.02-0''.04$ pixel scales. By checking the quality of the image obtained from the DJA , we ensure that they are eligible for further morphological studies , with $flag=2$. Then , we make $6.5'' \times 6.5$'' image cutouts for each source , based upon the background-subtracted images provided by the DJA , which are centered on their respective NIRCam detection coordinates. By using \texttt{PHOTUTILS} \citep{larry_bradley_2024_13989456} , we mask the surrounding area to suppress interference, keeping values higher than the 2$\sigma$ threshold in the exact center of the source. We perform this procedure on all of these six $JWST$ NIRCam bands (F150W, F200W, F277W, F356W, F410M, F444W). Both parametric and nonparametric morphological analyses are performed on these cutouts.  As for the Point Spread Function(PSF) , we use the ones provided by the DJA , which are generated using \texttt{PSFex}\citep{2013ascl.soft01001B} based on the custom selection methods MU\_MAX/MAG\_AUTO plane\citep{2007ApJS..172..219L} to yield better results than the default auto-selection from PSFEx.

\subsection{Parametric Morphology}

For parametric analysis , we focus primarily on the double Sérsic fitting , exploring the origins of the SMGs population by leveraging the properties of the bulge , specifically the bulge Sérsic index and the bulge-to-total luminosity ratio (B/T). The assumption of distinctions between the two evolution pathways is presented in Figure \ref{fig:merger_secular}.
We perform single- and double-component light profile fittings on our sample. Using the results of the single profile fitting as reference of the double sersic fitting and those of the previous studies , we set up the priors for the double profile fitting on this basis , to improve the accuracy and credibility of the results. 

\subsubsection{Single Sérsic fitting}
 
 We implemented our parametric modelings on the optical and near-infrared counterparts using \texttt{Pysersic} \citep{2023JOSS....8.5703P} , a Python-based code using Bayesian inference to fit the Sérsic profile to the galaxy light distributions. \texttt{Pysersic} is built on JAX \citep{jax2018github} for efficient numerical computation and utilizes NUMPYRO \citep{2019arXiv191211554P} for probabilistic inference. Compared to traditional methods such as \texttt{Galfit}\citep{2010AJ....139.2097P} , which rely on least squares fitting , \texttt{Pysersic} performs parameter estimations by sampling the posterior distributions , using the No U-Turn Sampler(NUTS) methods. There is no significant systematic difference between the results obtained by the two , which has been tested in recent studies on post-starburst galaxies\citep{2024ApJ...976...36Z} and on SMGs \citep{2025ApJ...982..200R}. 

We examine our results by comparing them with recent work on SMGs\citep{2025ApJ...982..200R}, including the sersic index and the physical size. The results of the single Sérsic modeling show overall good consistency , and details are shown in Figure \ref{fig:sersic-results} in the Appendix.

\subsubsection{Double Sérsic fitting}

In order to derive morphological information based on the bulges of the SMGs in more detail , we then perform the double Sérsic profiles on our samples , with the double Sérsic profile and a flat sky background fitted simultaneously. 

The Sérsic model profile fitted to both bulges and discs is:
\begin{equation}\label{eq1}
    I(R)=I_e\exp\{-b_n[(\frac{R}{R_e})^{1/n}-1]\}
\end{equation}
 where $R$ is the radius to the center of the galaxy , $R_e$(effective radius) serves as the radius that contains half of the total light/mass , $I_e$ is the intensity at that radius , $b_n$ is a constant solely dependent on n\citep{2004AJ....127.1917T} ,  and n measures the concentration of the overall light profile. When n=1 , the Sérsic profile reduces to an exponential profile that describes a pure disk , while n=4 corresponds to the de Vaucouleurs profile that describes an ellitpical galaxy. 

To ensure credible results of the double Sérsic fitting , we manually set the prior for the effective radius of both the bulge and the disc , as well as the Sérsic index $n\_bulge$ of the bulge. The effective radius for the bulge ,  \texttt{Re\_bulge},  follows a truncated Gaussian centered at 1 kpc , with a 50 $\%$ standard deviation and bounds from 0 to 2 kpc , while \texttt{Re\_disk} (the effective radius for the disc component) was centered at 4 kpc , with a truncated boundary from 2 to 12 kpc. The Sérsic $n\_bulge$ was assigned a loose Gaussian prior centered at 2.0(standard deviation = 1.0) to favor a bulge-like structure , while $n\_disk$(Sérsic index of the disc) and the prior of all the remaining parameters are estimated by the $autoprior()$ function, ensuring flexibility for disc-like components and robust fits across our galaxy sample. We then sequentially fit each band: F150W, F200W, F277W, F356W, F410M, and F444W. 

In the end , we obtain several parameters via Sersic modeling through \texttt{Pysersic}: the total integrated flux $flux$ , the position of Sérsic modeling $(x_c, \ y_c)$, the fraction of luminosity contained in the inner component(namely the bulge to total ratio) $f_1$ , the position angle $\theta$ , the ellipticy of both the bulge $e_1$ and the disc $e_2$ ,  the effective radius of both the bulge $r_1$ and disk $r_2$, the Sérsic index of bulge $n\_bulge$ and disc $n\_disk$. Due to various factors(such as low-quality imaging , contamination from neighboring sources , and incomplete data) , the total number of galaxies found in each band may decrease compared to the original sample.

\subsection{Non-Parametric Morphology}
In order to obtain more robust and comprehensive indicators , we then apply the nonparametric analysis to these samples using the \texttt{statmorph} \citep{2019MNRAS.483.4140R} package. The package focuses on the derivations of the CAS system\citep{2003ApJS..147....1C} , $Gini$ \& $M_{20}$\citep{2004AJ....128..163L} , and other important morphological indicators. \textbf{By checking the average per-pixel signal-to-noise ratio(SNR) within the Gini segmentation region for each aperature in the provided DJA catalogue , we confirm that the vast majority of the SMGs in our sample are eligible for the non-parametric measurement(SNR$\leq 3$) , with only $\leq 3$ SMGs with SNR$\leq3$ .} We then run this routine on both SMGs and field samples on band F277W because of the completeness of the image data compared to other bands , with the majority of them having a "good" fit results($Flag=0$). Several factors can be cited to explain the reason with results that are of poor quality($Flag\neq0$) , for example , artifacts , sources that are too faint and distorted , and the crowd environment causes the inefficiency of masking the surrounding sources. 
\subsubsection{CAS System}      

The CAS system is widely used in many classification scenarios. The C in the CAS system stands for the concentration index , which is defined as:
\begin{equation*} \label{eq5}
    C= 5\log{\frac{R_{80}}{R_{20}}}
\end{equation*}
where $R_{20}$ and $R_{80}$ represent the radius that contains 20 percent and 80 percent of the total luminosity , respectively.

A ， the asymmetry index , is calculated by subtracting the galaxy image by rotating $180^{\circ}$ from the original one; the exact mathematical expression can be written as:
\begin{equation*}\label{eq6}
    A = \frac{\sum_{i, j}|f_{i, j}- f_{i, j}^{180^{\circ}}|}{\sum_{i, j}|f_{i, j}|}-A_{bkg}
\end{equation*}
with $f_{i, j}$ and $f_{i, j}^{180^{\circ}}$ represents the flux of each individual pixel of the image and the one after rotating the image by $180^{\circ}$ , respectively. $A_{bkg}$ is the average background asymmetry.

The Smoothness(or Clumpiness) index , S , describes the smoothness of the galaxy light distribution. It is obtained by subtracting the original galaxy image with the smooth version of it which has already been convolved with a Gaussian kernel to smooth out the details with high spatial frequency and preserve the one with low spatial frequency. By applying this procedure , structures with high spatial frequency can be emphasized for further analysis. The equation has a similar form with the one above:
\begin{equation*}\label{eq7}
    S = \frac{\sum_{i, j}|f_{i, j}- f_{i, j}^s|}{\sum_{i, j}|f_{i, j}|}-S_{bkg}
\end{equation*}
where $f_{i, j}$ and $f_{i, j}^{180^{\circ}}$ represent the flux of each individual pixel of the image and the one after rotating the image by $180^{\circ}$ , respectively. $S_{bkg}$ is the smoothness of the background.

\begin{figure*}[htbp]
    \centering
    \includegraphics[width=1.0\linewidth]{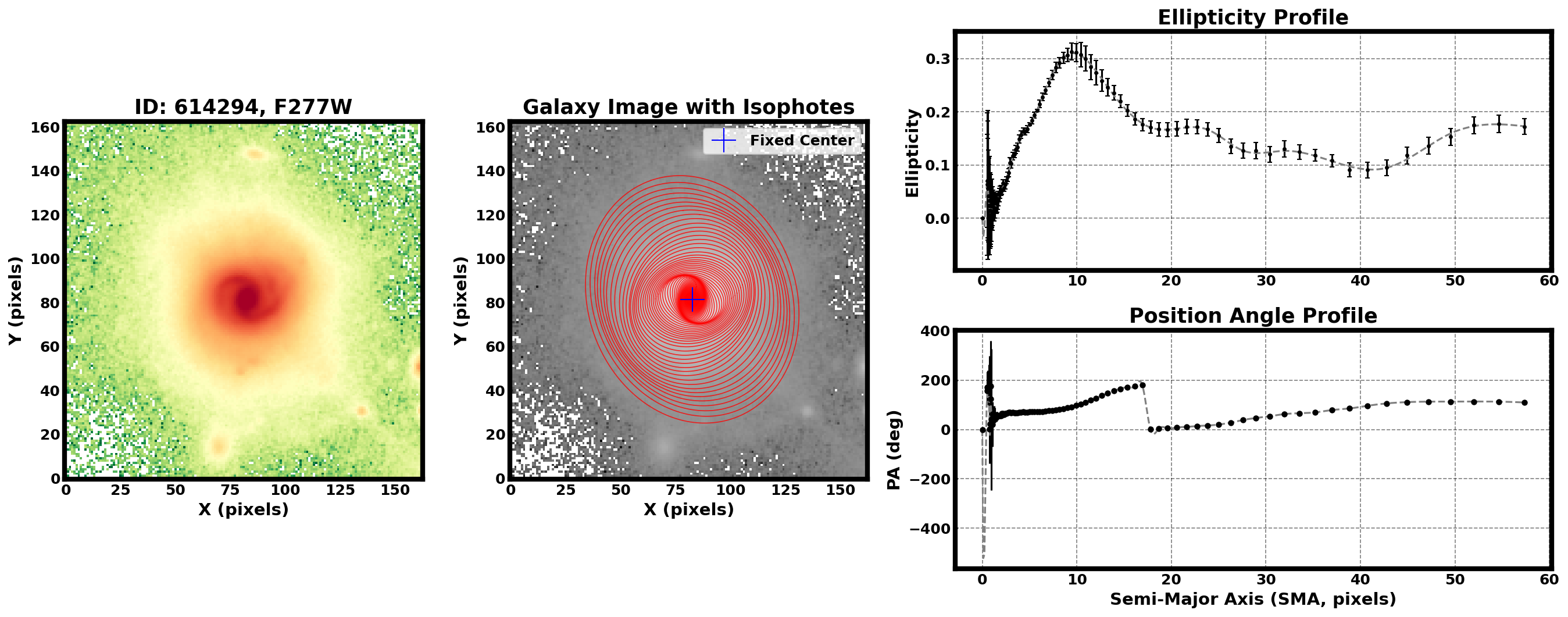}
    \caption{An example of bar identification via ellispe isophotal fitting of source ID: 614294 via ellipse-fitting techniques as an example. The blue cross marks the shared center of isophotes(red). }
    \label{fig:ellipse}
\end{figure*}

\subsubsection{Gini-$M_{20}$ Diagram}
One of the most commonly used merger classification methods is the $Gini-M_{20}$ diagram , which is used to select the one with potential merger activity from those non-mergers. To explain it concisely , the Gini coefficient was first introduced in the economics field to describe the inhomogeneous distribution of wealth across different classes. In galactic morphology , it measures the relative distribution of flux over pixels assigned to the particular galaxy , with $Gini=1$ corresponds to all of the light concentrated on one single pixel , while light spreads evenly across all pixels when $Gini=0$. The Gini coefficient of a galaxy with n pixels can be calculated by the following equation:
\begin{equation*}\label{eq8}
    G = \frac{1}{\bar{f} n(n-1)}\sum_{i=1}^{n}(2i-n-1)f_i
\end{equation*}
where $f_i$ is the flux of the i$th$ pixel , and $\bar{f}$ is the mean of the flux value.

The $M_{20}$ provides information regarding the normalized second-order moment of the brightest 20 percent of the galaxy's pixel , which is normalized by the total moment for all pixels. Galaxies with brighter , clumpy fragments have a larger value , whereas those with fewer fragments and a less dispersed light distribution have lower values. For galaxies with n pixel , the $M_{20}$ can be calculated by the following expression:

\begin{equation*}\label{eq9}
    M_{tot} = \sum_{i}^{n}M_i=\sum_{i}^nf_i[(x_i-x_c)^2+(y_i-y_c)^2]
\end{equation*}
in which $f_i$ stands for the flux measured at ith pixel , and the $(x_c, y_c)$ is the central position that minimize the $M_{20}$.

\subsection{Bar identification}
There are numerous approaches to identify a stellar bar within galaxies\citep{2010ApJS..186..427N, 2015A&A...582A..86H, 1991rc3..book.....D, 2007ApJ...657..790M, 2007ApJ...659.1176M, 2008ApJ...675.1141S, 2020ApJ...900..150Y, 2009MNRAS.393.1531G, 2005MNRAS.362.1319L, 2015ApJS..219....4S}, including visual inspection, two-dimensional decomposition , Fourier analysis , and ellipse isophotal fitting techniques. Each has its own costs and benefits. Several studies have been carried out to pinpoint the stellar bar within the high-z galaxies leveraging ellipse fitting since the advent of the JWST\citep{2024A&A...688A.158L, 2024MNRAS.530.1984L, 2023ApJ...945L..10G}. Taking all of the factors into account, we refer to the methods used in the previous research and follow them after tailored to our dataset. We employ three-step bar identification procedures , combining visual inspection , ellipse-fitting , and Sérsic fitting residuals.
\begin{enumerate}
    \item \textbf{Visual Inspection:}
    We visually check the cutouts to remove the one with high inclination and the one being too faint to be detected or contaminated by nearby sources, in which we perform an initial search for the candidates for stellar bar.
    \item \textbf{Ellipse Isophotal Fitting:}
    We used the photutils.isophote Ellipse package to identify the stellar bar in galaxy images via two-stage isophotal fitting process. We fit isophotes with free center$(x_0, y_0)$ , ellipticity$(\epsilon)$ , and the position angle(PA) , using an initial SMA of 5 pixels , $\epsilon=0.2$ , $PA=0$ , step\_size=0.1 , and max\_SMA=30 pixels , with median integration and 3-sigma clippings. The galaxy center was set as the mean of the inner isophote centers. We then fix the center and fit isophotes up to SMA=60 pixels with a step-size of 0.05 , ensuring the center stability($<10^{-6}$ pixel variation). We identified bar candidates by detecting the $\epsilon$($>0.1$) peaks in the profile , and visually compared them against bar-like features by comparing raw and isophote-overlaid images.  An example using the ellipse isophote fitting method is shown in Fig. \ref{fig:ellipse}.
    \item \textbf{Sérsic Fitting Residuals:}Last but not least ,  we examined the residuals of single Sérsic fittings to ensure the reliability of the identification results , which highlights the elongated structures consistent with bars provided additional evidence , ensuring robust classification.
\end{enumerate}

Bar candidates require either visual inspection or ellipse fitting confirmation to be labeled as 'tentative' but require both methods for confirmation. The final bar confirmation is determined by the presence of bar-like features in the Sérsic residuals. The residuals of the confirmed bar candidates via single Sérsic modellings are shown Fig \ref{fig:Bar} in the appendix.

\begin{table*}[htbp]
\centering
\begin{tabular}{lccccc}
\toprule
\textbf{Band (N)} & \textbf{$Re\_bulge$ [kpc]} & \textbf{$Re\_disk$ [kpc]} & \textbf{$n\_bulge$} & \textbf{$n\_disk$} & \textbf{B/T} \\
\midrule
F150W (N=116) & $1.18^{+0.26}_{-0.11}$ & $5.43^{+4.93}_{-1.85}$ & $2.00^{+1.04}_{-1.04}$ & $0.69^{+0.44}_{-0.44}$ & $0.09^{+0.26}_{-0.26}$ \\
F200W (N=100) & $1.19^{+0.51}_{-0.11}$ & $4.36^{+3.35}_{-1.15}$ & $1.90^{+1.09}_{-1.09}$ & $0.69^{+0.73}_{-0.73}$ & $0.22^{+0.31}_{-0.31}$ \\
F277W (N=125) & $1.45^{+0.51}_{-0.38}$ & $3.99^{+2.78}_{-0.89}$ & $1.47^{+1.85}_{-1.85}$ & $0.71^{+1.00}_{-1.00}$ & $0.46^{+0.31}_{-0.31}$ \\
F356W (N=106) & $1.42^{+0.60}_{-0.32}$ & $4.10^{+2.73}_{-1.30}$ & $1.33^{+1.81}_{-1.81}$ & $0.71^{+0.97}_{-0.97}$ & $0.53^{+0.31}_{-0.31}$ \\
F410M (N=99)  & $1.29^{+0.60}_{-0.22}$ & $4.23^{+2.88}_{-1.45}$ & $1.63^{+1.47}_{-1.47}$ & $0.71^{+0.93}_{-0.93}$ & $0.47^{+0.34}_{-0.34}$ \\
F444W (N=99)  & $1.26^{+0.69}_{-0.20}$ & $4.19^{+2.58}_{-1.41}$ & $1.48^{+1.51}_{-1.51}$ & $0.72^{+0.67}_{-0.67}$ & $0.58^{+0.33}_{-0.33}$ \\
\bottomrule
\end{tabular}
\caption{Summary of physical properties of SMGs across six JWST bands. The columns represent the effective radii of the bulge and disk ($r_\mathrm{eff,1}$ and $r_\mathrm{eff,2}$), their Sérsic indices ($n\_bulge$ and $n\_disk$), and the bulge-to-total luminosity ratio (B/T). }
\label{tab:physical_properties}
\end{table*}

\begin{table*}[!t]
\centering
\vspace{0.5em}
\begin{tabular}{lcccccccc}
\toprule
\textbf{Band} &
\textbf{Faint (\%)} &
\textbf{Bright (\%)} &
\textbf{\( B/T \) (F)} &
\textbf{\( B/T \) (B)} &
\textbf{\( p \)-val (B/T)} &
\textbf{\( n_{\mathrm{bulge}} \) (F)} &
\textbf{\( n_{\mathrm{bulge}} \) (B)} &
\textbf{\( p \)-val (\( n_{\mathrm{bulge}} \))} \\
\midrule
F150W & 53.3\% & 46.7\% & \(0.06 \pm 0.03\) & \(0.10 \pm 0.03\) & 0.384 & \(3.91 \pm 0.17\) & \(3.83 \pm 0.18\) & 0.057 \\
F200W & 55.7\% & 44.3\% & \(0.22 \pm 0.04\) & \(0.28 \pm 0.04\) & 0.686 & \(3.54 \pm 0.19\) & \(3.08 \pm 0.22\) & 0.974 \\
F277W & 50.9\% & 49.1\% & \(0.40 \pm 0.03\) & \(0.39 \pm 0.03\) & 0.929 & \(2.31 \pm 0.22\) & \(2.38 \pm 0.25\) & 0.652 \\
F356W & 54.0\% & 46.0\% & \(0.43 \pm 0.03\) & \(0.45 \pm 0.04\) & 0.922 & \(2.23 \pm 0.21\) & \(2.06 \pm 0.24\) & 0.769 \\
F410M & 53.6\% & 46.4\% & \(0.35 \pm 0.03\) & \(0.42 \pm 0.04\) & 0.905 & \(2.92 \pm 0.18\) & \(2.72 \pm 0.22\) & 0.899 \\
F444W & 53.6\% & 46.4\% & \(0.32 \pm 0.03\) & \(0.37 \pm 0.04\) & 0.459 & \(2.79 \pm 0.22\) & \(2.30 \pm 0.25\) & 0.510 \\
\bottomrule
\end{tabular}
\caption{KS test results for bulge-to-total ratio (\(B/T\)) and bulge Sérsic index (\(n_{\mathrm{bulge}}\)) between faint (``F'', SFR \(<175\ M_\odot\ \mathrm{yr}^{-1}\)) and bright (``B'', SFR \(\geq175\ M_\odot\ \mathrm{yr}^{-1}\)) SMGs across JWST bands. Percentages (cols.\ 2--3) give the fraction of sources in each subgroup. Values reported are medians \(\pm\) uncertainty (as you supplied). The \(p\)-values are from two-sample Kolmogorov--Smirnov tests; we treat distributions as indistinguishable when \(p>0.05\).}
\label{tab:combined_bt_n1_pvals}
\end{table*}

\begin{table*}[htbp]
\centering
\begin{tabular}{l c c c c c c}
\toprule
 & \multicolumn{3}{c}{\textbf{Bulge (\(Re\_bulge\)/ kpc)}} & \multicolumn{3}{c}{\textbf{Disk (\(Re\_disk\)/ kpc)}} \\
\cmidrule(lr){2-4} \cmidrule(lr){5-7}
Band & SFR $<$ 175 \(M_\odot~\text{yr}^{-1}\) & SFR \(\geq\) 175 \(M_\odot~\text{yr}^{-1}\) & P-value & SFR $<$ 175 \(M_\odot~\text{yr}^{-1}\) & SFR \(\geq\) 175 \(M_\odot~\text{yr}^{-1}\) & P-value \\
\midrule
F150W & 1.17 $\pm$ 0.05 & 1.11 $\pm$ 0.06 & 0.354 & 4.81 $\pm$ 0.40 & 6.28 $\pm$ 0.53 & 0.065 \\
F200W & 1.19 $\pm$ 0.06 & 1.14 $\pm$ 0.06 & 0.141 & 4.23 $\pm$ 0.38 & 5.05 $\pm$ 0.59 & 0.268 \\
F277W & 1.30 $\pm$ 0.06 & 1.50 $\pm$ 0.06 & 0.335 & 3.75 $\pm$ 0.29 & 4.43 $\pm$ 0.47 & 0.072 \\
F356W & 1.28 $\pm$ 0.07 & 1.47 $\pm$ 0.07 & 0.598 & 3.74 $\pm$ 0.35 & 4.96 $\pm$ 0.57 & \textbf{0.020} \\
F410M & 1.16 $\pm$ 0.06 & 1.34 $\pm$ 0.07 & 0.396 & 3.81 $\pm$ 0.37 & 4.61 $\pm$ 0.65 & \textbf{0.043} \\
F444W & 1.20 $\pm$ 0.07 & 1.35 $\pm$ 0.08 & 0.725 & 3.97 $\pm$ 0.36 & 4.62 $\pm$ 0.62 & \textbf{0.017} \\
\bottomrule
\end{tabular}
\caption{Effective radii for bulges (\(Re\_bulge\)) and disks (\(Re\_disk\)) across six bands, split by SFR (in \(M_\odot~\text{yr}^{-1}\)). Bold p-values ($<$ 0.05) show disks expand as the star formation rate rises.}
\label{tab:combined_sizes}
\end{table*}

\begin{table*}[htbp]
\centering
\begin{tabular}{lccccc}
\toprule
\textbf{Band} & \textbf{\( B/T \) Bin} & \multicolumn{2}{c}{Faint (\( SFR < 175 \, \mathrm{M_\odot~yr^{-1}} \))} & \multicolumn{2}{c}{Bright (\( SFR \geq 175 \, \mathrm{M_\odot~yr^{-1}} \))} \\
\cmidrule(lr){3-4} \cmidrule(lr){5-6}
& & \textbf{Fraction (\%)} & \textbf{\( n\_bulge \) (Median)} & \textbf{Fraction (\%)} & \textbf{\( n\_bulge \) (Median)} \\
\midrule
\multirow{3}{*}{F150W} & \( < 0.2 \) & 62.3\% (43/69) & \( 2.03^{+0.15}_{-0.54} \) & 63.8\% (30/47) & \( 2.19^{+0.03}_{-0.13} \) \\
 & 0.2--0.5 & 21.7\% (15/69) & \( 1.20^{+0.43}_{-0.46} \) & 23.4\% (11/47) & \( 1.07^{+0.68}_{-0.40} \) \\
 & \( > 0.5 \) & 15.9\% (11/69) & \( 1.25^{+0.45}_{-0.55} \) & 12.8\% (6/47) & \( 1.06^{+0.12}_{-0.33} \) \\
\midrule
\multirow{3}{*}{F200W} & \( < 0.2 \) & 44.3\% (27/61) & \( 2.13^{+0.05}_{-0.17} \) & 53.8\% (21/39) & \( 2.19^{+0.05}_{-0.17} \) \\
 & 0.2--0.5 & 29.5\% (18/61) & \( 0.76^{+0.75}_{-0.09} \) & 28.2\% (11/39) & \( 1.41^{+0.84}_{-0.72} \) \\
 & \( > 0.5 \) & 26.2\% (16/61) & \( 0.73^{+0.70}_{-0.06} \) & 17.9\% (7/39) & \( 0.75^{+0.30}_{-0.08} \) \\
\midrule
\multirow{3}{*}{F277W} & \( < 0.2 \) & 30.6\% (22/72) & \( 2.15^{+0.13}_{-0.37} \) & 22.6\% (12/53) & \( 2.22^{+0.05}_{-0.01} \) \\
 & 0.2--0.5 & 27.8\% (20/72) & \( 1.28^{+3.00}_{-0.58} \) & 28.3\% (15/53) & \( 1.47^{+1.47}_{-0.80} \) \\
 & \( > 0.5 \) & 41.7\% (30/72) & \( 0.75^{+1.14}_{-0.09} \) & 49.1\% (26/53) & \( 0.76^{+1.71}_{-0.10} \) \\
\midrule
\multirow{3}{*}{F356W} & \( < 0.2 \) & 23.8\% (15/63) & \( 2.23^{+1.12}_{-0.06} \) & 20.9\% (9/43) & \( 2.21^{+1.55}_{-0.02} \) \\
 & 0.2--0.5 & 19.0\% (12/63) & \( 3.01^{+1.79}_{-2.02} \) & 25.6\% (11/43) & \( 1.61^{+4.29}_{-0.94} \) \\
 & \( > 0.5 \) & 57.1\% (36/63) & \( 0.72^{+1.20}_{-0.06} \) & 53.5\% (23/43) & \( 0.69^{+0.33}_{-0.04} \) \\
\midrule
\multirow{3}{*}{F410M} & \( < 0.2 \) & 28.8\% (17/59) & \( 2.20^{+0.10}_{-0.09} \) & 25.0\% (10/40) & \( 2.25^{+0.24}_{-0.03} \) \\
 & 0.2--0.5 & 20.3\% (12/59) & \( 2.57^{+1.40}_{-1.79} \) & 30.0\% (12/40) & \( 2.54^{+1.78}_{-1.84} \) \\
 & \( > 0.5 \) & 50.8\% (30/59) & \( 0.68^{+0.61}_{-0.02} \) & 45.0\% (18/40) & \( 0.70^{+0.35}_{-0.04} \) \\
\midrule
\multirow{3}{*}{F444W} & \( < 0.2 \) & 28.8\% (17/59) & \( 2.21^{+1.00}_{-0.10} \) & 20.0\% (8/40) & \( 2.24^{+1.14}_{-0.02} \) \\
 & 0.2--0.5 & 18.6\% (11/59) & \( 3.22^{+1.54}_{-0.84} \) & 15.0\% (6/40) & \( 4.35^{+0.48}_{-1.22} \) \\
 & \( > 0.5 \) & 52.5\% (31/59) & \( 0.67^{+0.85}_{-0.01} \) & 65.0\% (26/40) & \( 0.68^{+0.24}_{-0.03} \) \\
\bottomrule
\end{tabular}
\caption{Distribution of bulge Sérsic index \( n\_bulge \) in \( B/T \) bins for faint (\( SFR < 175 \, \mathrm{M_\odot~yr^{-1}} \)) and bright (\( SFR \geq 175 \, \mathrm{M_\odot~yr^{-1}} \)) SMGs in different JWST bands. Fractions in each bin are shown along with the median \( n\_bulge \) and asymmetric errors.}
\label{tab:bt_n1_sfr}
\end{table*}

\begin{table*}[htbp]
\centering
\begin{tabular}{lccccc}
\toprule
\textbf{Bulge Type (N)} & \textbf{Stellar Mass} & \textbf{SFR} & \textbf{B/T} & \textbf{Sérsic $n$} & \textbf{$R_\mathrm{eff}$} \\
 & $(10^{11} M_\odot)$ & $(M_\odot~\text{yr}^{-1})$ & & & (kpc) \\
\midrule
Classical Bulge (N=27)    & $1.6^{+1.1}_{-0.9}$  & $162.2^{+200.9}_{-73.1}$ & $0.3^{+0.2}_{-0.1}$ & $4.6^{+1.1}_{-1.9}$ & $2.0^{+0.0}_{-0.5}$ \\
Major Merger Bulge (N=4)  & $1.3^{+1.3}_{-0.9}$  & $152.5^{+273.6}_{-116.6}$ & $0.6^{+0.1}_{-0.0}$ & $4.4^{+1.0}_{-0.2}$ & $1.4^{+0.7}_{-0.7}$ \\
Clump Sinking Bulge (N=16)& $1.6^{+0.6}_{-0.4}$  & $204.4^{+294.2}_{-95.5}$ & $0.5^{+0.1}_{-0.1}$ & $2.7^{+0.6}_{-0.3}$ & $2.0^{+0.1}_{-0.6}$ \\
Pseudo Bulge (N=21)       & $0.8^{+0.4}_{-0.4}$  & $151.4^{+137.1}_{-122.5}$ & $0.1^{+0.1}_{-0.1}$ & $1.4^{+0.8}_{-0.6}$ & $1.1^{+0.6}_{-0.1}$ \\
Unclassified Bulge (N=47) & $0.9^{+0.9}_{-0.6}$  & $128.8^{+153.3}_{-52.0}$ & $0.7^{+0.2}_{-0.2}$ & $0.7^{+0.2}_{-0.0}$ & $1.6^{+0.3}_{-0.3}$ \\
\bottomrule
\end{tabular}
\caption{Median of the physical properties by bulge type, with the lower and upper bounds corresponding to the 25th and 75th percentiles. All values are rounded to one decimal place.}
\label{tab:bulge_properties}
\end{table*}

\section{RESULTS}
\subsection{Morphological parameters across bands}
\begin{figure}[]
    \centering
    \includegraphics[width=1\linewidth]{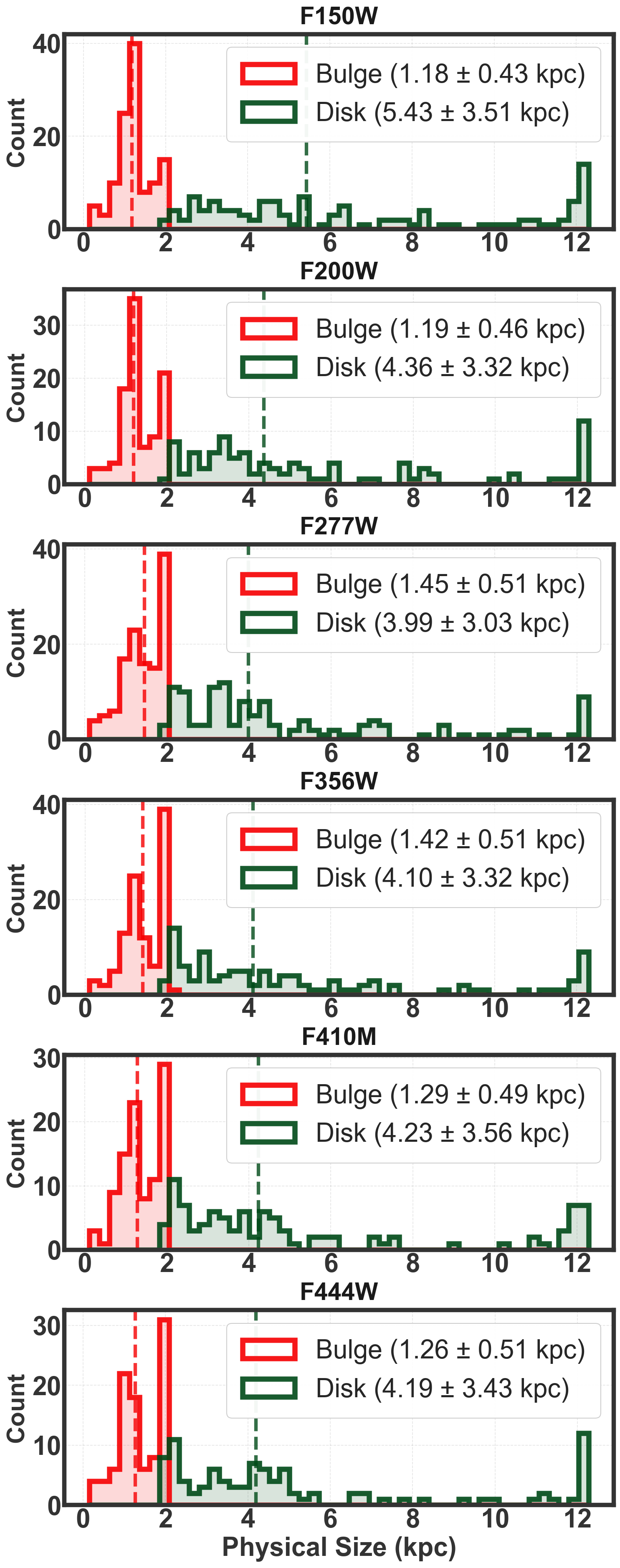}
    \caption{Distribution of the effective radius of bulge(\texttt{Re\_bulge}) and disk(\texttt{Re\_disk}) across F150W, F200W, F277W, F356W, F410M, and F444W, with the red and dark blue dotted vertical line indicating the median of the effective radius of the bulge and the disk respectively.}
    \label{fig:effective_radii}
\end{figure}

\
We measure the half-light radius and Sérsic indices(including bulges and discs) of SMGs across the six bands. We use the $r\_{hat}\approx1.0$ to assess the convergence of the Markov Chain Monte Carlo (MCMC) measurements , making sure that all markov chains are all converged to same underlying distribution of the results. The overall results of the properties is shown in Table \ref{tab:physical_properties}, and an intuitive plot about the effective radii of the bulge and the disk is displayed in Fig \ref{fig:effective_radii}.  To make it concise , we show the median values of effective radius and Sérsic indices of band F150W and F444W, with median values of half-light radius $R_{bulge}^{150W}=1.18 \pm 0.41 \ kpc$ , $R_{disc}^{F150W}=5.74\pm3.65 \ kpc$ , $R_{bulge}^{F444W}=1.35\pm0.51 \ kpc$ , $R_{disc}^{F444W} = 4.19\pm 3.58 \ kpc $ , and median values of Sérsic indices 
$n_{bulge}^{F150W}=2.03\pm0.75$ , $n_{disc}^{F150W}=0.68\pm 0.42$ , $n_{bulge}^{F444W}=0.98\pm1.52$ , and $n_{disc}^{F444W}=0.72\pm0.71$. The overall results of the physical size show values comparable with the optical size of the rest frame and the infrared size of the rest frame measured by \cite{2025ApJ...988..135M} ,  \cite{2025ApJ...982..200R} , \cite{2024A&A...691A.299G} , and \cite{2025ApJ...979..229M}. Several factors can be considered when it comes to the offset between the results and the one from previous works. One of the most notable causes is that single Sérsic modeling considers the light distribution of the bulge and disc as a whole , making the half-light radius larger than the one from the bulge and slightly smaller than the one of the disc derived from double Sérsic modeling , which is also proved by one of the first JWST works \citep{2022ApJ...939L...7C}. Factors such as sample selection criterion and approaches used to measure the light distribution can be taken into account to explain the discrepancy.

In addition to physical size and light distribution index , we also measure the bulge-to-total luminosity ratio(B/T) , which presents the fraction of total luminosity contributed by the bugle component of the galaxy. For simplicity, we only show the median value of B/T of band F150W and F444W , with $B/T^{F150W}=0.09\pm0.26$ and $B/T^{F444W} =0.58\pm0.33$. 

To ensure the credibility of our results , we perform the accuracy test on our results of single Sérsic and double Sérsic fitting in F277W. The accuracy of the single Sérsic shows $97.3\%^{+1.53}_{-6.46}$ for the n , and $98.4\%^{+1.0}_{-3.5}$ for the r\_eff.  The accuracy of the double Sérsic shows that $91.2\%^{+6.5}_{-60.1}$ for the f\_1 , $81.8\%^{+13.8}_{-9.5}$ for the n\_bulge, $93.9\%^{+5.3}_{-15.2}$ for n\_disk, $87.2\%^{+11.0 }_{-14.1}$ for Re\_bulge, $92.3\%^{+5.8}_{-11.4}$ for Re\_disk.

\begin{figure}[]
    \centering
    \includegraphics[width=1\linewidth]{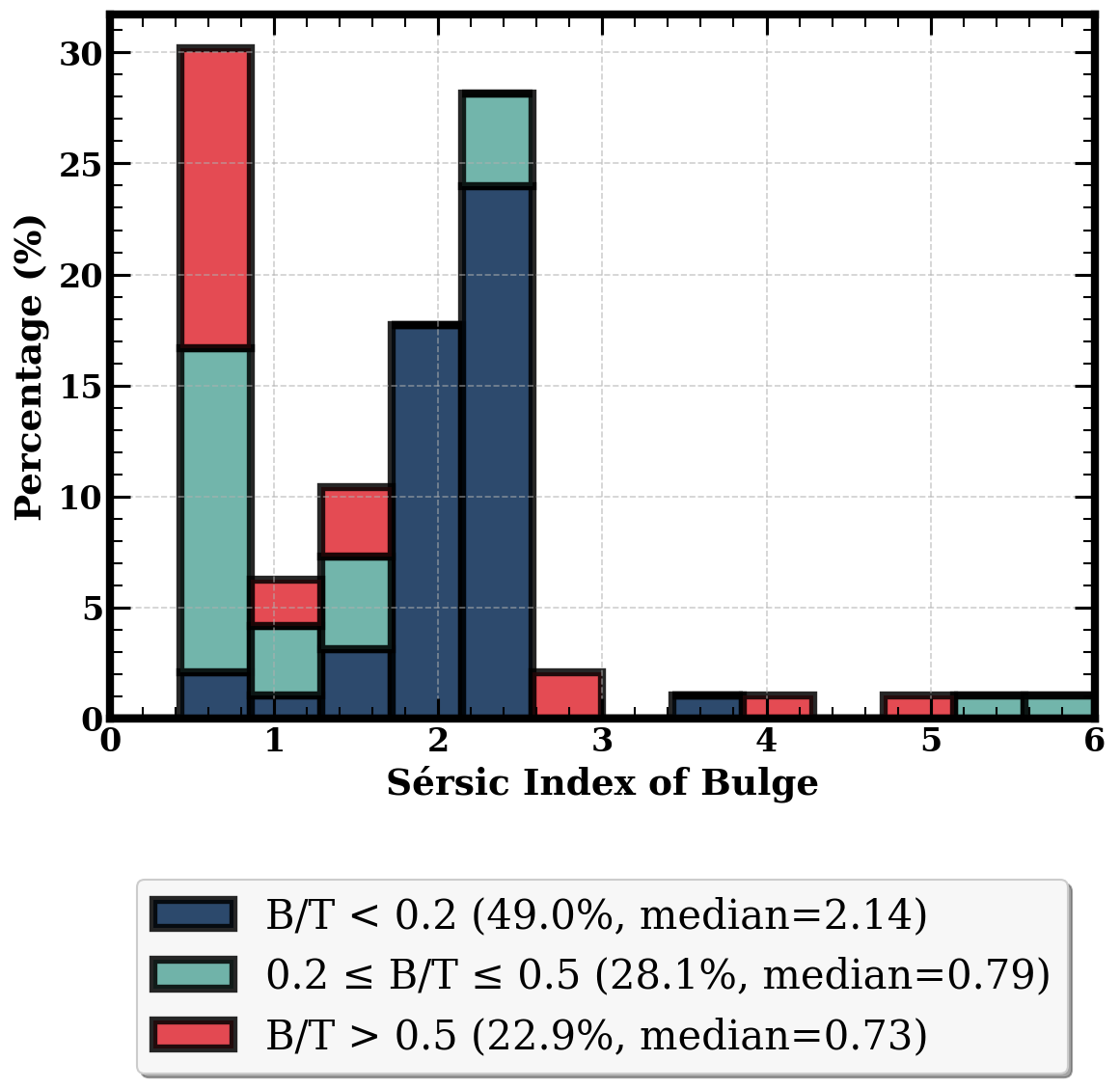}
    \caption{The distribution of bulge Sérsic indices across three B/T bins($B/T<0.2$ , $0.2\leq B/T \leq0.5$ , $B/T>0.5$) in band F200W.}
    \label{fig:F200W_bt_sersic}
\end{figure}

\subsubsection{SFR dependent Morphology}
To establish the idea whether the bright and faint population of SMGs is subjected to the same distribution , we divide our samples into two bins: one with \texttt{SFR} $< 175 \ M_{\odot}~yr^{-1}$ , and the other with \texttt{SFR} $> 175 \ M_{\odot}.yr^{-1}$. We then performed the Kolmogorov-Smirnov (KS) test on the frequency distribution of B/T and Sérsic index of the bulge between these two groups.  Table \ref{tab:combined_bt_n1_pvals} summarizes results for all bands.

In general , there are no significant statistical differences across the board. In particular , we highlight the F444W due to its longest wavelength , which enables better penetration through dust , providing a clearer view of the stellar structures within SMGs. In F444W , bright SMGs exhibit a B/T of $0.60\pm0.05$ and $n\_bulge$ of $0.91\pm0.27$ , while the fiant ones show a B/T of $0.52\pm0.04$ and $n\_bulge$ of $1.58\pm0.18$ , with \texttt{p-value} $>0.05$ showing no significant distinctions between these two groups.  Although $p<0.05$ in the F150W , we find no significant statistical difference between these two groups by checking the cdf diagram between the groups. 

In addition , we perform the KS test on the effective radii of the bulge component and the disk component. The results are shown in Table \ref{tab:combined_sizes} , which shows larger radii in both the bulge and the disk of the bright population comparing it to the faint ones across the board. The \texttt{p-values} in the F356W, F410M, and F444W show an even greater distinction between the two groups.

\subsubsection{Structural trend by B/T}

To explore the evolving structural trend of the bulges within the center of the SMGs , we group them into three B/T bins: $B/T < 0.2$ , $0.2 < B/T < 0.5$ , and $B/T > 0.5$ , based upon previous splits by \texttt{SFR}. We discard the results with poor quality, by setting a mask to rule out the one with the negative flux(fitting results for each band). Detailed information on the B/T splits can be found in Table \ref{tab:bt_n1_sfr}(with a more intuitive plot of the distribution of bulge Sérsic indices divided by three B/T bins in Appendix \ref{fig:BT_sersic_6_band}). Intriguingly , it turns out that there is an apparent trend within the distribution. SMGs with higher B/T bin lean towards lower bulge's lower Sérsic index , whilst the ones with lower B/T sit right at the position with higher Sérsic index(with an example in band F200W shown in Fig \ref{fig:F200W_bt_sersic}).

For the faint SMGs , the low $B/T$ bin ($B/T < 0.2$) leads in shorter-wavelength bands , capturing $62.3\%$ (43/69) in F150W and $44.3\%$ (27/61) in F200W , with $n\_bulge = 2.03\text{--}2.23~(\sigma = 0.36\text{--}1.30)$. These high $n\_bulge$ values reflect steep , compact bulges. The intermediate bin ($0.2 < B/T < 0.5$ , $18.6\text{--}29.5\%$) shows transitional profiles with $n\_bulge = 0.76\text{--}3.22~(\sigma = 1.58\text{--}2.31)$ , while the high $B/T$ bin ($15.9\text{--}57.1\%$) dominates in longer wavelengths , peaking at $57.1\%$ (36/63) in F356W and $52.5\%$ (31/59) in F444W , with $n\_bulge = 0.67\text{--}1.25~(\sigma = 0.98\text{--}2.06)$ , indicating flatter profiles.

For the bright SMGs , the whole picture shifts slightly. The low $B/T$ bin remains prominent in F150W ($63.8\%$ , 30/47) and F200W ($53.8\%$ , 21/39), with $n\_bulge = 2.19\text{--}2.25~(\sigma = 0.32\text{--}0.76)$ , mirroring the faint SMGs' compact bulges. The intermediate bin ($15.0\text{--}30.0\%$) yields $n\_bulge = 1.07\text{--}4.35~(\sigma = 0.98\text{--}2.81)$ , but the high $B/T$ bin surges in longer wavelengths , reaching $65.0\%$ (25/40) in F444W with $n\_bulge = 0.68~(\sigma = 1.49)$ , compared to only $12.8\%$ (6/47) in F150W with $n\_bulge = 1.06~(\sigma = 0.67)$.

We also have to consider the 

In brief , the bright SMGs maintain similar Sérsic indices of the bulge , while still having higher B/T than the faint SMGs. All of these patterns highlight a structural gradient tied to both wavelength and flux.

\subsubsection{Properties of the SMGs in different bulge categories}
We highlight the properties of our sample of SMGs categorized by different types of bulges in Table \ref{tab:bulge_properties}. Classical bulges exhibit relatively high stellar mass ($1.6^{+1.1}_{-0.9} \times 10^{11}~M_\odot$) and elevated star formation rates ($162.2^{+200.9}_{-73.1}~M_\odot~\text{yr}^{-1}$). In particular, they show some of the broadest diversity in structural properties , with a bulge-to-total ratio (B/T) of $0.3^{+0.2}_{-0.1}$, a Sérsic index of $n=4.6^{+1.1}_{-1.9}$, and an effective radius of $R_\mathrm{eff}=2.0^{+0.0}_{-0.5}$~kpc.

Within this category, bulges built through major mergers are sites of particularly extreme cases. These systems have high stellar masses ($1.3^{+1.3}_{-0.9} \times 10^{11}~M_\odot$) and SFRs ($152.5^{+273.6}_{-116.6}~M_\odot~\text{yr}^{-1}$) , along with the highest median B/T ($0.6^{+0.1}_{-0.0}$) , a high Sérsic index ($n=4.4^{+1.0}_{-0.2}$) , and compact bulges ($R_\mathrm{eff}=1.4^{+0.7}_{-0.7}$~kpc).

Clump-sinking bulges have stellar masses similar to average classical bulges ($1.6^{+0.6}_{-0.4} \times 10^{11}~M_\odot$) and even higher SFRs ($204.4^{+294.2}_{-95.5}~M_\odot~\text{yr}^{-1}$). Structurally , they lie between merger-built bulges and pseudobulges: their B/T is $0.5^{+0.1}_{-0.1}$ , Sérsic index is $2.7^{+0.6}_{-0.3}$ , and effective radius is $2.0^{+0.1}_{-0.6}$~kpc.

Pseudobulge-hosting SMGs exhibit the lowest median stellar mass among the classified categories ($0.8^{+0.4}_{-0.4} \times 10^{11}~M_\odot$) and slightly lower SFRs ($151.4^{+137.1}_{-122.5}~M_\odot~\text{yr}^{-1}$). Morphologically , they have the lowest B/T ($0.1^{+0.1}_{-0.1}$) , lowest Sérsic indices ($n=1.4^{+0.8}_{-0.6}$) , and relatively compact bulges ($R_\mathrm{eff}=1.1^{+0.6}_{-0.1}$~kpc). 

Finally , SMGs with unclassified bulges show moderate stellar masses ($0.9^{+0.9}_{-0.6} \times 10^{11}~M_\odot$) and the lowest SFRs ($128.8^{+153.3}_{-52.0}~M_\odot~\text{yr}^{-1}$). They exhibit the lowest Sérsic indices ($n=0.7^{+0.2}_{-0.0}$) , intermediate B/T ($0.7^{+0.2}_{-0.2}$) , and modest bulge sizes ($R_\mathrm{eff}=1.6^{+0.3}_{-0.3}$~kpc).

\subsection{Non-parametric metrics}
Beyond Sérsic fitting, we examine the morphology of SMGs using the CAS sysstems($C , A, S$) , and the $Gini-M_{20}$ metrics across F150W-F444W , leveraging F277W as the benchmark for non-parmetric analysis , since the parametric fitting cannot fully disclose the morphological information , especially for clumpy discs and SMGs with merger signatures.  We follow the identical categorization above based on \texttt{SFR} , on the basis of which we present the distribution of the non-parametric metrics. We checked the results and ensured the results robust estimations(\texttt{flag = 0} , yielding 101 SMGs with high-quality results. An detailed table with non-parametric metrics is shown in the Appendix Fig \ref{fig:non_parametric_metrics}.

\begin{figure*}
    \centering
    \includegraphics[width=1.04 \linewidth]{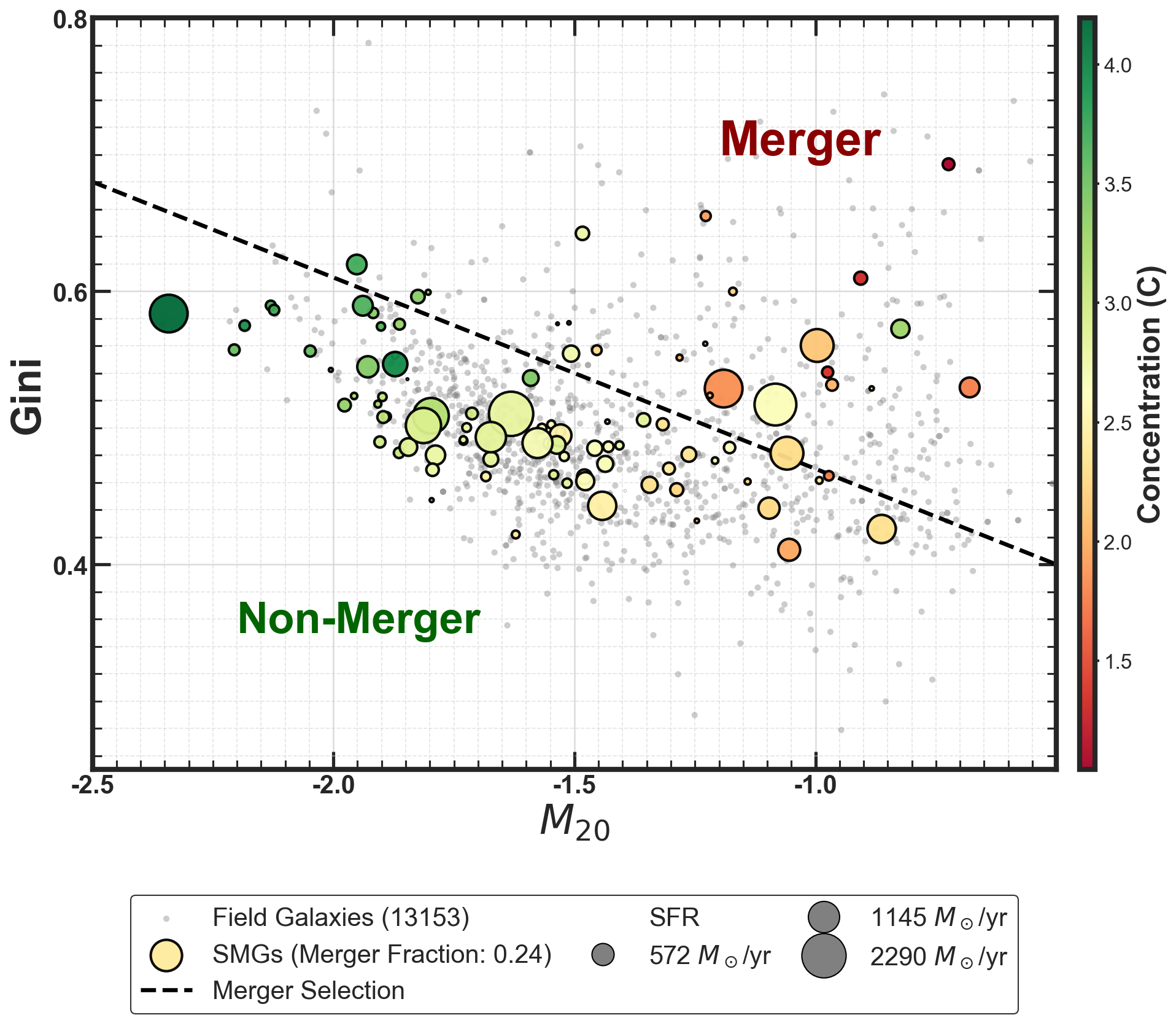}
    \caption{\texttt{Gini-M20} relation for both the SMGs and the field galaxies in the PRIMER-COSMOS field , with the dashed line indicating the boundry of the galaxies with merger signatures and those non-mergers , with detailed description in \cite{2008ApJ...672..177L}.}
    \label{fig:Gini-M20}
\end{figure*}

For CAS systems, faint SMGs(61/101) show a median concentration $C = 2.78$ and asymmetry $A =0.18$, painting a picture of compact , smooth , nearly featureless disks. However , bright SMGs(40/101) shift to lower concentration but higher asymmetry , with $C = 2.75$ and $A = 0.21$ , suggesting a more disturbed inner structure than their fainter counterparts. We then perform the KS test and the result indicates that there is no significant difference between these two groups despite the marginal difference in concentration and asymmetry , suggesting an overall uniformity across the 2mJy threshold in these measures.
 
To probe the origin of the SMGs derived from morphological information , we then inspect the Gini-M20 plot, which is shown in Fig \ref{fig:Gini-M20}. The overall merger rate of our sample SMGs is $24\%$. For fiant SMGs , the median is $Gini=0.51$ and $M_{20}=-1.55$, while the median is $Gini=0.50$ and $M_{20}=-1.50$ in bright SMGs. KS test again suggests similarity (with $p=0.67$ for Gini , $p=0.49$ for $M_{20}$) , with most SMGs clustering in the disk region. Still, $24\%$(24/101) falls into the merger zone(27.5\% bright , 8/40 , and 21.3\% faint , 13/61) , consistent with the minor dynamical disturbance. Additionally , the $M_{20}$ ，which tracks the degree of separation of distribution of luminous clumps within the boundaries of the galaxies , shows better separation in distinguishing the bright population from the faint ones than all the other non-parametric indices including Gini. This phenomenon has been confirmed in \cite{2019AAS...23312805N}. Our sample have a broad redshift range , spanning from $z\sim1-6$ , which corresponds to rest-frame wavelength range of $\sim0.4-1.39  \mu m$ in band F277W , with the median $0.84 \mu m$ (16th-84th percentile: $0.57-1.12 \mu m$) , spanning from optical to near-IR regime . Despite previous studies showing the variations in the rest-frame wavelength may affect the measurements of $M_{20}$ \citep{2025ApJ...982..200R} , we find no significant statistical differences by performing the KS test (p-value$=0.19$) on the $M_{20}$ value between the two groups divided by $0.75 \mu m$ . Additionally , the offset between the two groups is modest ($\Delta M_{20} = 0.07$) , with median $M_{20}= -1.48$ for the $<0.75 \mu m$ group and median $M_{20}$ for the $>0.75 \mu m$ group. Consequently , the impacts due to variations of the rest-frame wavelength are limited in 
this work.

The overall non-parametric metrics echo the results of parametric light distribution modellings ,  reinforcing the idea of prevalence of disky structures and internally driven scenarios in the SMGs population from previous studies.

\begin{figure*}
    \centering
    \includegraphics[width=1.0\linewidth]{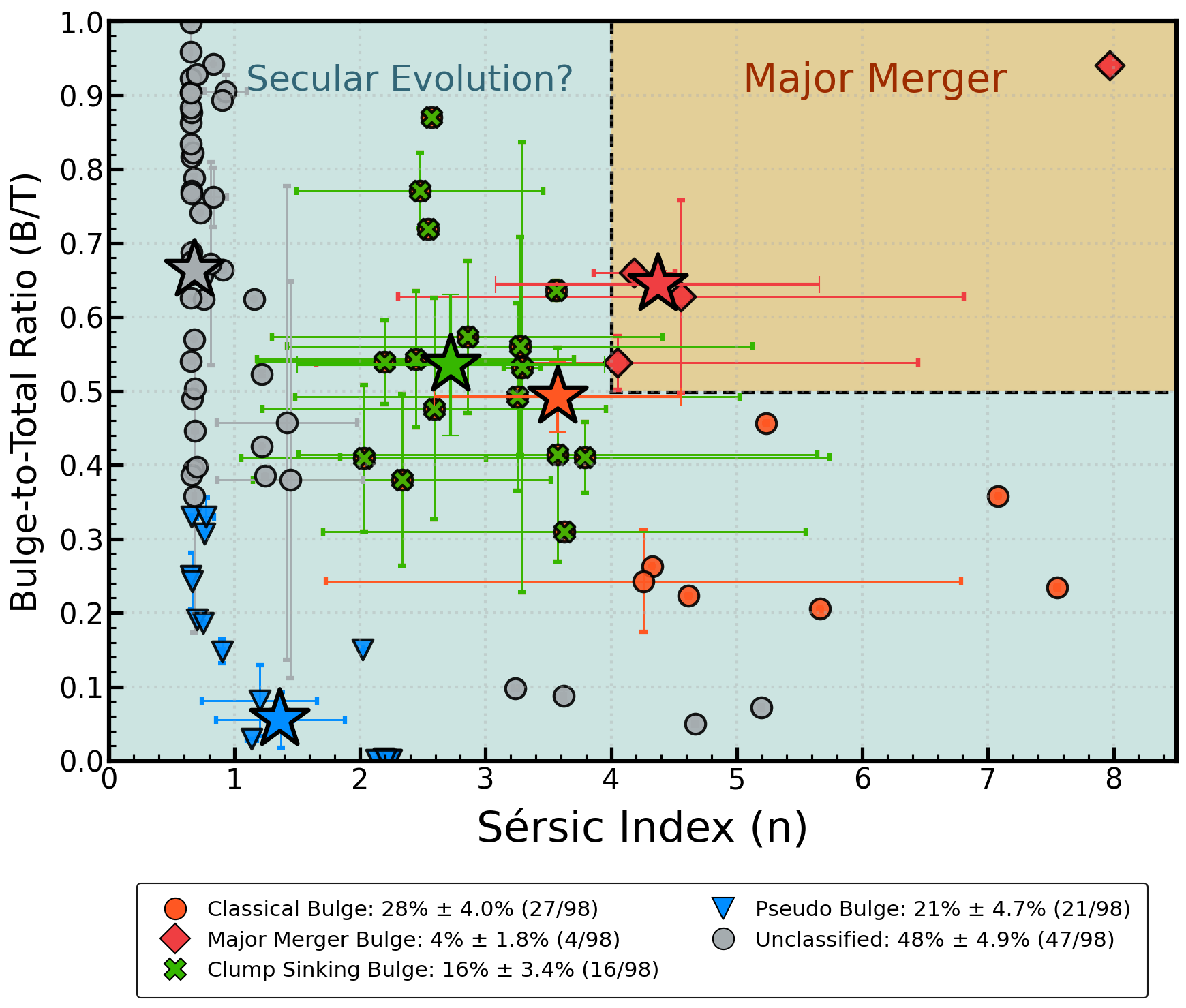}
    \caption{The distribution of different types of bulges based on the  classification scheme using sérsic index and bulge-to-total luminosity ratio on band F277W after the removal of the strongly disturbed SMGs , with the big star symbols showing the median of each group.}
    \label{fig:bt_vs_sersic}
\end{figure*}

\section{Discussion}
\subsection{Pseudo-bulge or Classical Bulges?}
First of all , before diving into detailed discussions , we must clarify the definition of the classical bulges and pseudobulges , which are sometimes quite ambiguous to elucidate. In general , they are distinguished by their assumed formation mechanisms: pseudo-bulges are thought to arise from gas inflows triggered by their bars or spiral arms within the disc plane , or any internal processes , whereas the classical bulges are often attributed to the violent relaxation induced by either the local disk instability or the major mergers. However , this theoretical divide crumbles under scrutiny , as observational diagnostics to confirm such origins remains scarce , especially at high redshift.

For example , several early simulations show that massive star-forming clumps in gas-rich , high-redshift discs can merge into the central structures resembling dispersion-dominated classical bulges owing to dynamical frictions , with $\alpha$-enhanced abundances and the $R^{1/4}$ light distribution\citep{2004A&A...413..547I, 2009ASPC..419...23E, 2000MNRAS.312..194N} , while some others show that the majority of clumps at high-redshift , gas-rich disks are short-lived and gravitationally unbound ,  which dissipate before entering the center of the galaxy , making it unclear to fully explain it being the sole origin of the classical bulge at high redshift , gas-rich galaxies\citep{2020MNRAS.494.1263M, 2017MNRAS.465..952O}. More interestingly , \cite{2022MNRAS.514.2497K} , using FLAMINGO simulation ,  points out that classical bulge dominates in early time , while pseudobulge is more prevalent in later epoch.

On the observational side , previous work on the 'cosmic grape'\citep{2024arXiv240218543F} reveals the fact that numerous star-forming clumps are formed through the disk instabilities with weak feedback effects , suggesting in situ formation of clumps in high redshift , gas-rich , rotating disks. Another study shows that there is a statistically significant correlation between clumpiness and the bulge-to-disc flux ratio\citep{2024ApJ...960...25K} , strongly suggesting that bulges are evolutionarily related to clumps. In addition , some models suggest such bulge should have lower Sérsic index and significant rotation, correlating to the bar\citep{2012MNRAS.427..968H, 2016A&A...588A..42S, 2012MNRAS.422.1902I}. Yet , the conventionally accepted notion of the formation of classical bulges is bound to major merger activity , and these clump-built structures need further confirmation , despite being observationally similar.

We set our boundaries for classical bulge and pseudobulge based on the list of the classcial-pseudobulge classification criteria , provided in John Kormendy article\citep{2016ASSL..418..431K, 2004ARA&A..42..603K, 2012ApJS..198....2K}, in which contains detail distinctions between these two structures , including the morphological properties , abundance of metallicity , and velocity field etc. We only take the standards regarding the Sérsic index and bulge-to-total luminosity 
ratio($B/T$) , owing to the limitation of our data. The followings are our standards:

\begin{itemize}
    \item Classical Bulge(General Classfication):
    \begin{itemize}
    \item   $n\geq2$
    \item   $ B/T> 0.2$(Optional, for more restricted purpose)
    \end{itemize}
    \item Classical Bulge(Merger-built):
    \begin{itemize}
    \item $n\geq 4$
    \item $ B/T> 0.5$
    \end{itemize}
    \item  Classical Bulge(Clump sinking):
    \begin{itemize}
    \item $2<n<4$
    \item $B/T>0.2$(Optional, for more restricted purpose)
    \end{itemize}
    \item Pseudobulge(General Classification):
    \begin{itemize} 
    \item  $n<2$
    \item    $B/T \leq 0.35$
    \end{itemize}

\end{itemize}
and please keep in mind that each of these classification criteria has its failure rate ranging from $0\%$ to  $20\%$. To make it clear , there is no absolute standard in these two types of structure and the classification scheme is for rough reference only.

We then use the standard above to gauge the number of each category in the F277W band due to the completeness of the image data , and it turns out to be quite intriguing. The results are shown in Fig. \ref{fig:bt_vs_sersic}. After removing the SMGs with strongly disturbed morphology(majority of the removals are identified as 'merger' and 'clumpy' , but not necessarily all of them) by visual inspection , the number of general classification of classical bulge yields 40 out of 98 , and  27 out of 98 with a more rigorous version. For the pseudobulge , the classification shows 21 out of 98 bulges that are qualified for the selection criterion. Additionally , there are only four merger-built classical bulges , whereas 36 of the bulges are certified as the classical bulge built via clumps sinking(16 out of 98 for more rigorous purpose). Interestingly , half of the bulges($47/98$) fail to fall into any of our categories , with the vast majority clustering around the area where the higher B/T and lower Sérsic coexist , suggesting that the majority of the bulge components in the higher redshift SMGs are still in a phase of growth and becoming more and more prominent. 

We have mentioned in the previous section that SMGs in our sample tend to have higher $B/T$  and lower Sérsic index at the same time , while lower $B/T$ and higher Sérsic index simultaneously , which contradicts the paradigm that a classical bulge must have higher B/T plus higher Sérsic index and a pseudo-bulge must have lower B/T and lower Sérsic simultaneously.  The p-values shown in Table \ref{tab:combined_bt_n1_pvals} suggest no significant statistical difference between the two SFR splits in all morphological parameters. In addition , there is a decreasing trend in the Sérsic index in both $< \ 175 M_\odot~yr^{-1}$ and $\geq \ 175 M_\odot~yr^{-1}$ group , which is more evident in the bright population. This suggests the fact that the bulges in the SMGs have an extended , disk-shaped geometry , similar to local pseudobulges in terms of Sérsic index but having a much larger $B/T$ value. The trend confirms that the bulge component is more prominent in the longer wavelength , revealing a wavelength-dependent effect possibly tied to dust penetration.

The physical sizes of both bulges and disks in all six of these JWST bands show no significant statistical difference by implementing the KS test, although the physical sizes of the bright group are slightly larger than their fainter counterparts(e.g., median $R_e \sim 6.28 kpc$(bright) vs. $4.81 \ kpc$(faint) in band F115W). This compactness challenges the notion that classical bulges , if merger-induced , should expand markedly while still dominating the luminosity of the galaxies. Instead , the prevalence of pseudobulge , clump-sinking bulges , plus unclassified bulge with high B/T ($\sim 90\%$) and the modest merger fraction($24\%$) point toward a secular evolution scenario , where gas-rich discs at high redshift foster in situ bulge formation through violent disk instabilities\citep{2009Natur.457..451D}.  On top of that , we have mentioned in the previous section that there is a subgroup with high Sérsic index but lower B/T , in other words , the small classical bulge.  This type of bulge is considered to be the product of minor mergers or clumps migration\citep{2016ASSL..418..431K, 2012MNRAS.421..333S, 2016ASSL..418..199G, 2016ASSL..418..413C} , some of which in the local universe are also embedded in the pseudobulge driven by bar instability\citep{2015MNRAS.446.4039E} , making the whole a composite bulge. In addition , \cite{2024ApJ...974L..28B} shows that bulge formation starts prior to the quiescence phase , imposing new constraints on the evolution pathways of the high-z star-forming galaxies. 

However, \cite{2024Natur.636...69T} points out that the in-situ formations of spheriods in submillimeter-bright galaxies (SMGs) are induced by mergers, leveraging high-resolution ALMA observations of 146 bright SMGs($S_{870\mu m} = 7.8  \text{mJy}$ , $\log(M_*/M_\odot) = 11.0$ , $ \text{SFR} = 624  M_\odot  \text{yr}^{-1}$) in COSMOS and GOODS-S fields and being supported by simulations. Morphologically, they show bulge-like profiles (median Sérsic index ( $n \approx 1.6$ ), spergel index ($ v = -0.26 $)) and trixal shape($ q = 0.71$, $C/A = 0.53$). Combined to our sample(which is less intensive SMGs,  merger-rate being 27.5\% in bright group, and 21.3\% in faint group via non parametric methods), still, secular evolution takes precedence over major mergers, albeit merger being more prominent in the formation of brigher, more intense SMGs.

Despite the theoretical and observational constraints above , we must clarify the possibility that these outcomes could be merely mathematical manifestations of the model without substantial physical interpretation. For instance , half of the SMGs have a bulge with high B/T($\sim 0.7$) but low Sersic($\sim1$) , with B/T accuracy spanning $91.2\%^{+6.5}_{-60.1}$ , which may denote that B/T is unstable, indicating that the single sérsic profile suffices to describe the light distribution of the SMGs at high redshift , with the possibility of the absence of bulges inside the SMGs at high-z , or the nucleus is still in the process of formation.

With all regarded , the traditional dichotomy of pseudo-bulge and classical bulge shows limitations under the influence of gas dynamics and dust at high redshift , rendering SMGs as a possible hybrid class required further observational constraints.

\subsection{Origin of SMGs: Merger-induced or Internally Driven?}
As shown in Figure \ref{fig:Gini-M20}, we show the Gini-$M_{20}$ plot of the SMGs sample in F277W , split by two flux bins following the identical configuration as the one above. It is evident that the majority of SMGs(90/101) lie in the non-merger regionIn either the faint or bright group , SMGs with higher star formation rate(as marked by the size of the scatter point) tend to reside in the area of late-type galaxies , which is basically composed of disc and irregular galaxies , suggesting that the intense starburst activities occur in the non-merging phase (or pre-merging phase, if the major merger origin is validated , referring to the evolution pathway of Gini and M20 in this hydrodynamical simulation \cite{2019A&A...632A..98C} for details) , consistent with starburst systems driven by internal dynamics\citep{2014PhR...541...45C}.

Besides that , our nonparametric metrics in previous section paint a different picture from the major-merger-induced scenario. In F277W , the Gini and $M_{20}$ remain relatively stable across flux bins. , with p-value($>0.05$) indicating no significant structural shifts. The concentration($C\sim 2.75$) and asymmetry($A\sim0.20$) indices further underscore settled , less disturbed disk-like systems , consistent with those intense starburst systems in \cite{2014ARA&A..52..291C}. The morphological outcomes align with multiple previous studies. \cite{2025ApJ...978..165H} , using high resolution ALMA 870 $\mu m$ imaging , finds out that SMGs at $z\sim2-3$ exhibits disk-like feature($n=1$) though measurements of compact star forming regions($R_e=0.6 \ kpc$) as well as clumpy sub-structures supporting an in-situ star formation via violent disk instabilities. Similarly , \cite{2018ApJ...861....7F} demonstrates that dusty star-forming galaxies with far-IR luminosities($L_{IR} =10^{11}-10^{13}L_\odot$) present disk-like morphology and compact far-IR emission areas($R_{e,FIR}\sim \ 1kpc$) , indicating starbursts in gas-rich disks.  

\cite{2024A&A...688A..53L} provide a new dynamical framework for dusty star-forming galaxies(DSFGs) , in which lopsidedness in 64 \% of the DSFGs is primary driven by asymmetric gas accretions and secondarily driven by minor mergers. Despite some impacts on lopsidedness , major merger is just a case-by-case basis and shows no statistical significance. Additionally , the study finds galaxies with high lopsidedness residing in isolated environment , suggesting it being an internal dynamical process.  Accretion-induced lopsidedness facilitates violent disk instability, forming clumps that migrate inward to build diffuse extended disk-like bulge secularly while maintaining high SFRs\citep{2009Natur.457..451D}. This accounts for our low asymmetry and high concentration , as the VDI maintains disk-like morphology. 

\begin{figure}
    \centering
    \includegraphics[width=1\linewidth]{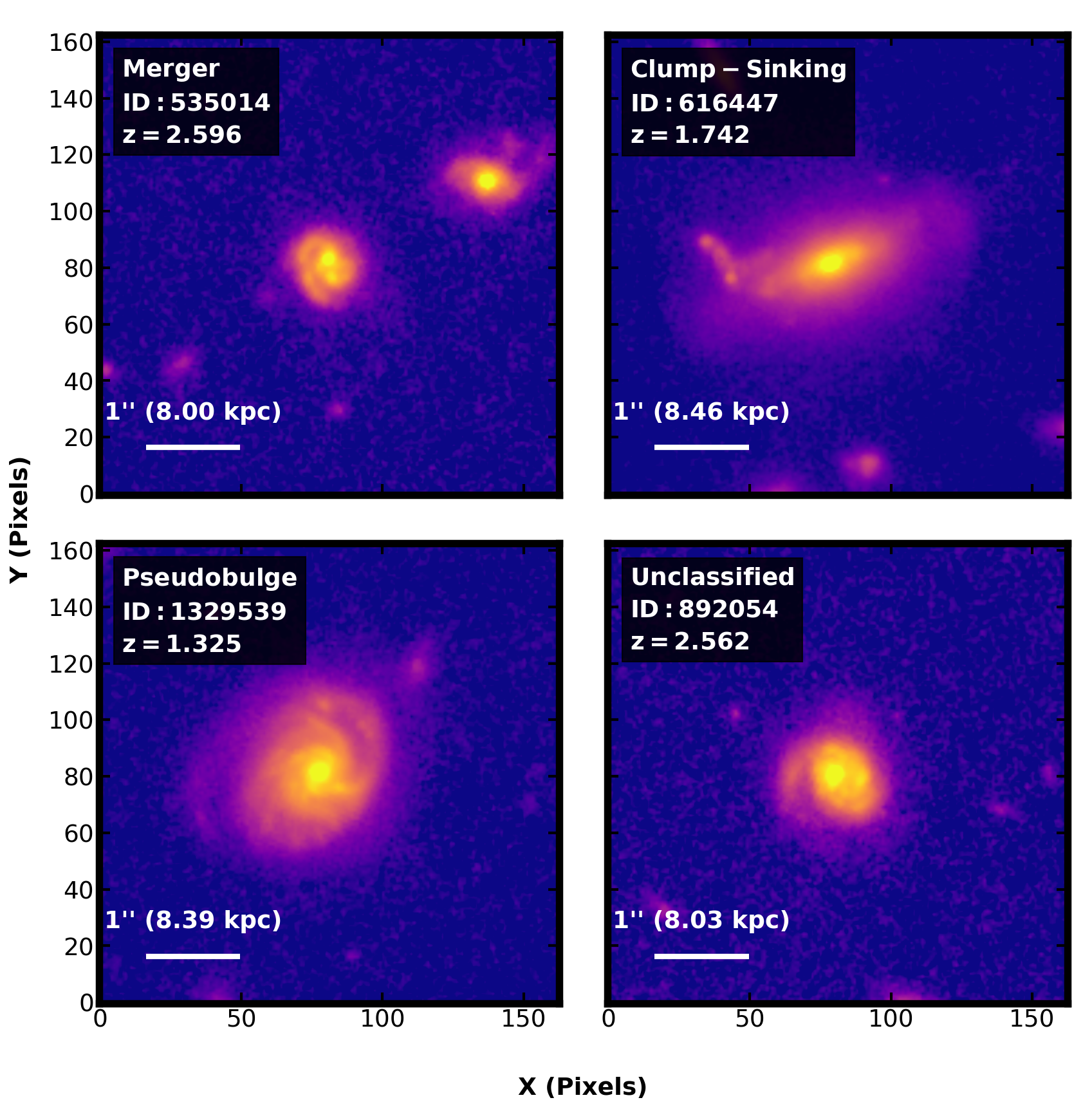}
    \caption{Example of SMGs with different bulge type($165''\times165''$ , including merger-built bulge , clump-sinking bulge , pseudobulge , and unclassified bulge, with the white horizontal line in each individual plot stands for the scale of 1 arcsec (0.03''/pixel).}
    \label{fig:placeholder}
\end{figure}

Only $24\%$(24/101) of the SMGs fall in the merger region , with bright($27.5\%$) groups having slightly higher merger-rate than the faint group($21.3\%$) , a fraction which is too modest to drive the bulk of our sample's star formation. This aligns with the \cite{2018ApJ...861....7F} merger fraction in ULIRGs and the \cite{2025ApJ...978..165H} $\sim 16 \ \%$ merger signatures , indicating an enhancements in star formation from mergers in luminous SMGs. \cite{2024A&A...688A..53L} notes that the minor merger boosts the lopsidedness , triggering disk instabilities\citep{2024A&A...691A.299G} and gas inflows to the center and forming a compact core , which aligns with our finding of small classical bulges.

The data points towards an internally driven scenario , where gas-rich discs at high redshift sustain starbursts through violent disk instability rather than extreme merger activities\citep{2009Natur.457..451D}. Clumps identified in our SMGs are consistent with predictions from the models of gravitationally unstable, turbulent discs\citep{2010MNRAS.407.1223R, 2014MNRAS.442.1230R}. The transient , "pre-merger" stage inadequately explains our SMGs' persistent disk-like feature and high star formation rate throughout the non-merger phase , if the merger-origin doctrine is valid. Instead , the observed clumps and gas compression may arise in situ , fueled by the filamentary inflows , with mergers playing a secondary role , which is decent in the sense that a denser , more gas-rich environment in the early universe can yield the starburst activity with comparable star formation rate like the local ULIRGs, which are typically associated with major mergers\citep{2012ApJ...753L..37D, 2021A&A...651A..42P, 2021A&A...646A.101P, 2022A&A...662A..94P}, but through smaller scale internal dynamical process. The predicted evolution of the galaxy population in  The internal structure, such as the bulge , disk , and stellar bar , can be fabricated through internal gas compressions and manifest the intrinsic spin of the dark matter halo , which is supported by several recent simulations\citep{2021ApJ...919..135D, 2024A&A...689A.293M, 2025A&A...697A.236L}. The observations of the JWST reveal modest evidences against the idea of formation of SMGs induced by major merger , pointing to the fact that the formation mechanisms of these high-z dusty star-forming systems lean heavily on internal dynamics , with external interaction enhancing , rather than initiating , their evolutionary pathway.

\subsection{Visual Morphology and Stellar Bar}
Thanks to the advent of the James Webb Space Telescope(JWST) , we are now able to visually inspect hidden details behind those dusty facades of the SMGs , with unprecedented high sensitivity and spatial resolution. Still , we perform the visual inspections on the SMGs in band F277W due to the completeness of the image data , with the rest-frame wavelength spanning from the optical to near-infrared regime($0.84_{-0.27}^{+0.28}~\mu m$).  We find that $\sim34\%$ of the SMGs exhibit clear undisturbed disk-like morphology and $\sim22\%$ with clumpy feature. The majority of the remaining SMGs has smooth , flatten morphology, except for the minority displaying the merger signatures($\sim20\%$).  Within the disk group , $\sim66\%$ features apparent spiral arms and $\sim 43\%$ has stellar bar(5 confirmed bar + 13 tentative candidates).  We then visually check the residuals of the Sérsic fittings to ensure credibility and 
reliability of bar detection , producing 9 (9/43 , $\sim21\%$ within the disk-like group) SMGs with a strong , confident bar-like structure that facilitate gas accretion funneling into the center of the galaxy, with the residuals shown in Appendix \ref{fig:Bar}. All of the bar candidates are all clustered at $z\sim1-3$ , peaking at $z\sim2$. Yet , the fraction of the total bar candidates could be underestimated because of the high inclination and the faint detections.
Interestingly , we notice that several of our edge-on SMGs have traces of filamentary gas accretion , and the other two exhibit point source features , indicating potential AGN activities.

The stellar bar plays a significant role in galaxy evolution , such as driving the internal processes , triggering gas inflows , forming boxy/peanut-shaped structures , influencing AGN fuelings , indicating evolutionary stage , redistributing baryonic matters , etc. \cite{2025ApJ...979..229M} shows $8\%\pm3\%$ of the SMGs with confirmed bar-like features, which is in agreement with our findings($\sim 7 \%$). All things considered, it is highly probable that the formations of structures such as stellar bars, spiral arms , and  bulges are in situ , caused by internal gas compressions, rather than external interferences.
    
\section{Conclusions}

In this study , we leverage the high resolution and precision of the James Webbs Space Telescope(JWST) NIRCam imaging to probe the morphology of 125 $z\sim1-6$ SMGs in the \texttt{PRIMER-COSMOS field}.All of our SMGs have robust detections of ALMA , with all of the physical properties presented in this paper retrieved from the A3COSMOS catalogue \citep{2024A&A...685A...1A}. The followings are our primary findings:
\begin{itemize}
    \item Secular Evolution as primary mechanisms:
    Morphological analysis of 125 ALMA-detected submillimeter galaxies(SMGs) in the PRIMER-COSMOS field , using the JWST/NIRCam imaging(F150W, F200W, F277W, F356W, F410M, F444W) , shows $76\%$ of the SMGs with a disk-like structure , which is potentially  driven by secular evolution via violent disk instabilities and filamentary gas inflows , which challenge the major merger-dominated paradigm.
    \item SFR-dependent morphology:
    Besides being intrinsically different at the stellar mass and the star formation rate , double sérsic modellings reveal bright SMGS have higher bulge-to-total luminosity ratio($B/T=0.60\pm0.05$) and lower bulge Sérsic index($n\_bulge = 0.91 \pm 0.27$) than the fainter one( $B/T = 0.52 ± 0.04$ , $n\_bulge = 1.58\pm 0.18$) in band F444W ,  suggesting varying degrees of internal evolution and different formation pathway.
    \item Disk-Dominated Structures:
    Non-parametric metrics in band F277W for 101 SMGs yield low asymmetry and the higher concentration($A=0.19_{-0.07}^{+0.09}$ , $C= 2.77_{-0.38}^{+0.40}$) , with the majority of our samples falling in the non-merger region , aligning with that of disk-dominated starburst group. Only 24\% shows merger signatures , consistent with $16 - 21\%$ merger rate from \cite{2024A&A...691A.299G} , \cite{2018ApJ...861....7F} , and \cite{2025ApJ...978..165H}.
    \item Bulge formation and Hybrid-origin Model:
    Bulge classification yields 16\% of SMGs with clump sinking bulge($2<n\_bulge<4$, $B/T>0.2$) and 4\% merger-built bulge. 48\% (47/98) of our sample have unclassified bulges , with the vast majority showing high B/T ($\sim 0.7$) but low Sérsic indices ($n\_bulge \sim 0–1.2$) , suggesting ongoing VDI-driven bulge formation. This supports a secular evolution model: 79\% secular , 21\% minor merger-induced lopsidedness \citep{2024A&A...688A..53L}.
    \item Visual Morphology :
    JWST/NIRCam imaging shows that 34\% of 125 $z\sim1-6$ SMGs in the PRIMER-COSMOS field have disk-like structures along with spiral patterns and stellar bars that are either confirmed or tentative (5 confirmed , 13 tentative). Clumpy features in 22\% of the SMGs indicate violent disk instability (VDI)-driven star formation while 20\% show signatures of mergers and the remaining objects exhibit smooth flattened profiles. Sérsic fitting residuals reveal 9 SMGs in 21 percent of disk systems that exhibit strong barred structures potentially driving starburst activity at their centers. Several edge-on SMGs display filaments of gas which confirms VDI yet two other SMGs contain point-like features that point to AGN activity.
\end{itemize}

Our work demonstrates the dominance of the internally driven secular evolution framework for the formation pathways of the SMGs , while mergers play a secondary role. However , stellar bar , or any other substructures of the inner part of the galaxies have not been ruled out while performing double Sérsic light distribution modellings , which requires further scrutinizations. More precise , in-depth studies via spectroscopic methodologies with ALMA and JWST will be crucial to probe the kinematics nature , constrain clump masses , and confirm AGN activity , shedding light on the dynamics of these high-z , dusty starburst systems.


\section*{Acknowledgements}
We appreciate comments from the anonymous reviewer which improve the quality of this work. We also extend our gratitude to Luis C. Ho and Daizhong Liu for providing insightful advice that expands the scope and depth of this work. 

This work is based on observations made with the NASA/ESA/CSA James Webb Space Telescope. The data were obtained from the Mikulski Archive for Space Telescopes at the Space Telescope Science Institute , which is operated by the Association of Universities for Research in Astronomy , Inc., under NASA contract NAS 5-03127 for JWST. These observations are associated with program 1837. The data described here may be obtained from the MAST archive at \dataset[doi:10.17909/dhrq-3k05]{https://doi.org/10.17909/dhrq-3k05}.

 Y.A. acknowledges the support from the National Natural Science Foundation of China (NSFC grants 12173089), the Strategic Priority Research Program of the Chinese Academy of Sciences (Grant No.XDB0800301) , and the National Key R\&D Program of China (2023YFA1608204). This work is based on observations made with the NASA/ESA/CSA James Webb Space Telescope. The data products presented herein were retrieved from the Dawn JWST Archive (DJA). DJA is an initiative of the Cosmic Dawn Center (DAWN), which is funded by the Danish National Research Foundation under grant DNRF140. 

\software{Astropy\citep{2013A&A...558A..33A, 2018AJ....156..123A, 2022ApJ...935..167A} , Matplotlib\citep{Hunter:2007} , Photutils\citep{larry_bradley_2024_13989456} , Pysersic\citep{2023JOSS....8.5703P} , Statmorph\citep{2019MNRAS.483.4140R} , Topcat\citep{2017arXiv171101885T}, Grizli\citep{2023zndo...8370018B}}

\facilities{JWST, ALMA}

\clearpage

\appendix
\section{Distribution of the bulge Sersic indices by B/T bins and single sersic , and Non-parametric results divided by star formation rate}

\begin{figure}[hbp]
    \centering
    \includegraphics[width=1\linewidth]{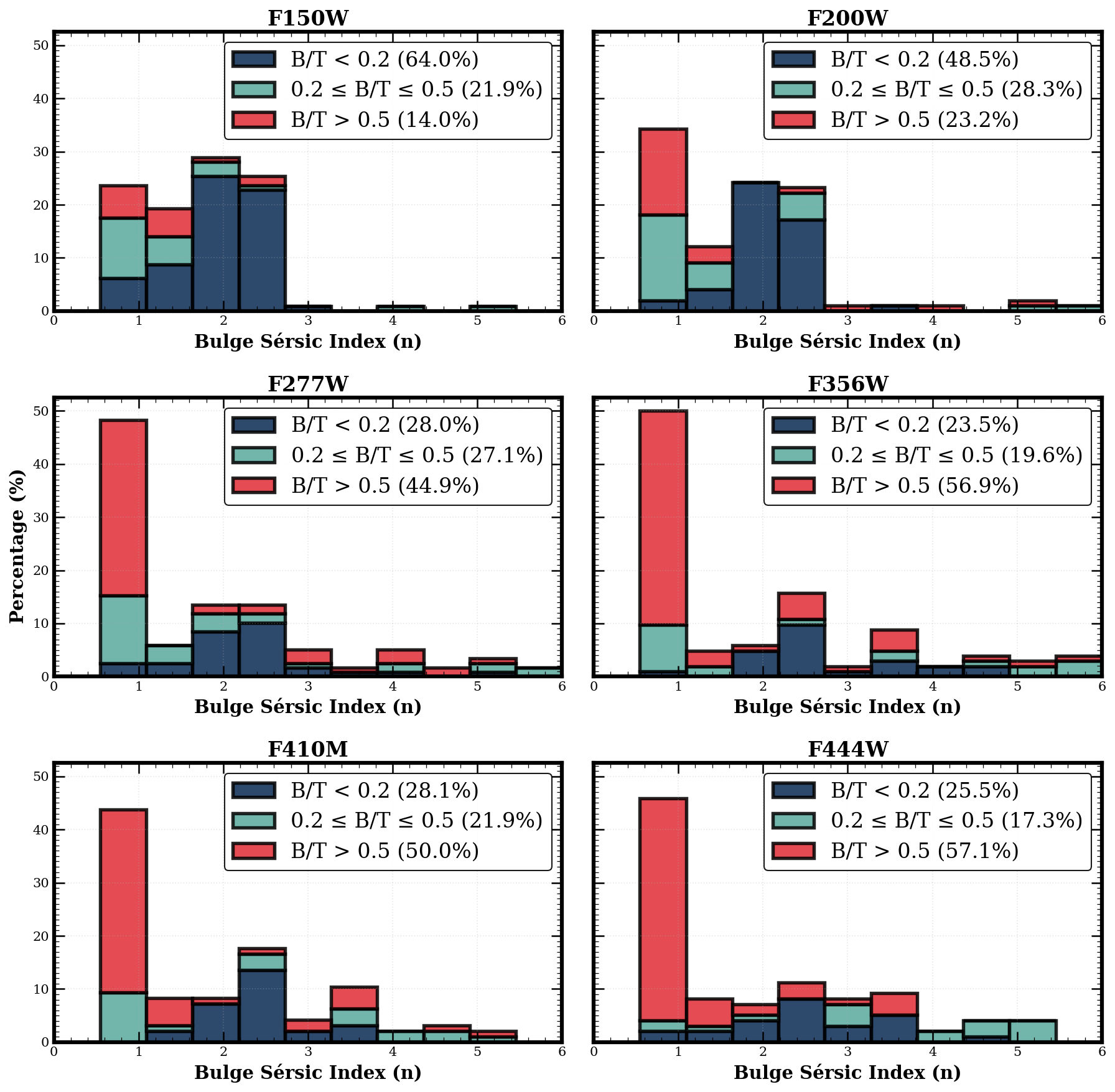}
    \caption{The distribution of the bulge Sérsic index across six band(F150W, F200W, F277W, F356W, F410M, F444W) divided by three B/T bins($B/T<0.2$, $0.2<B/T<0.5$, $B/T >0.5$).}
    \label{fig:BT_sersic_6_band}
\end{figure}
\begin{figure}[hbp]
    \centering
    \includegraphics[width=0.9\linewidth]{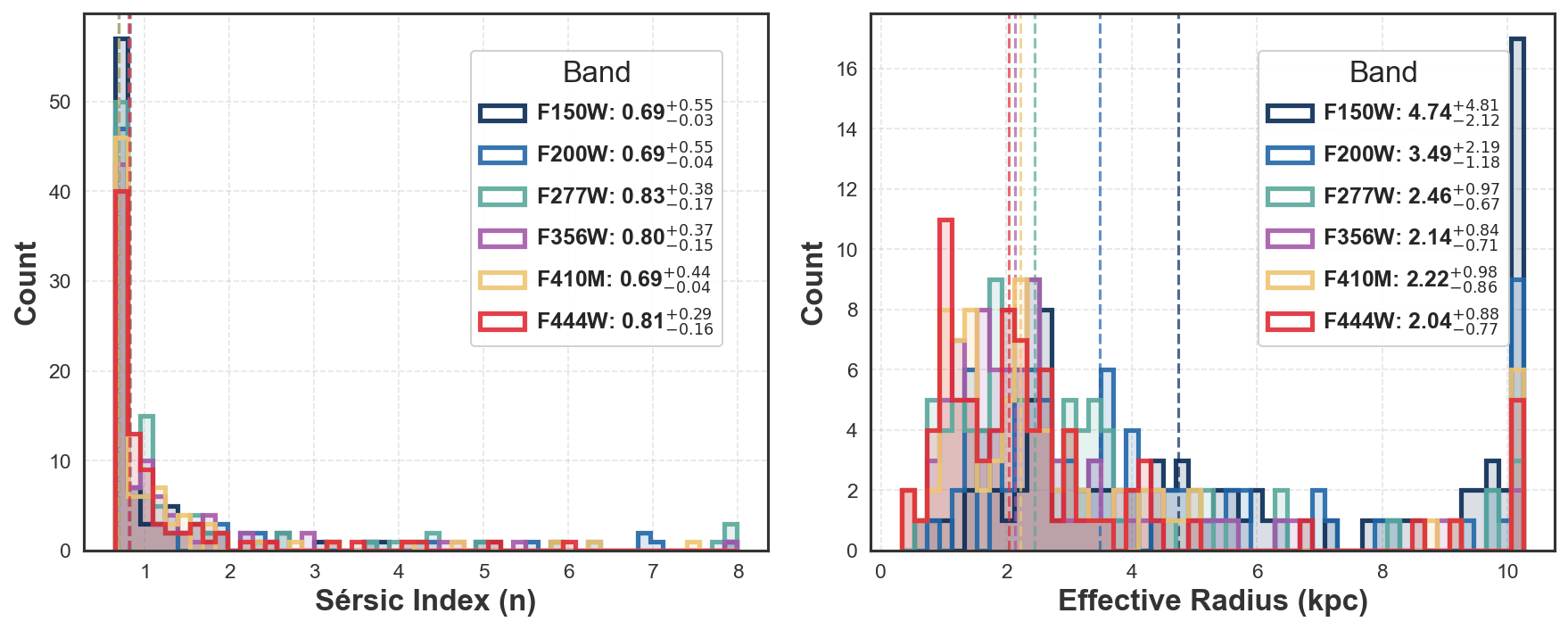}
    \caption{Resuls of the single Sérsic fitting: Distribution of Sérsic index and effective radius for JWST bands.}
    \label{fig:sersic-results}
\end{figure}

\begin{figure}[hbp]
    \centering
    \includegraphics[width=0.9\linewidth]{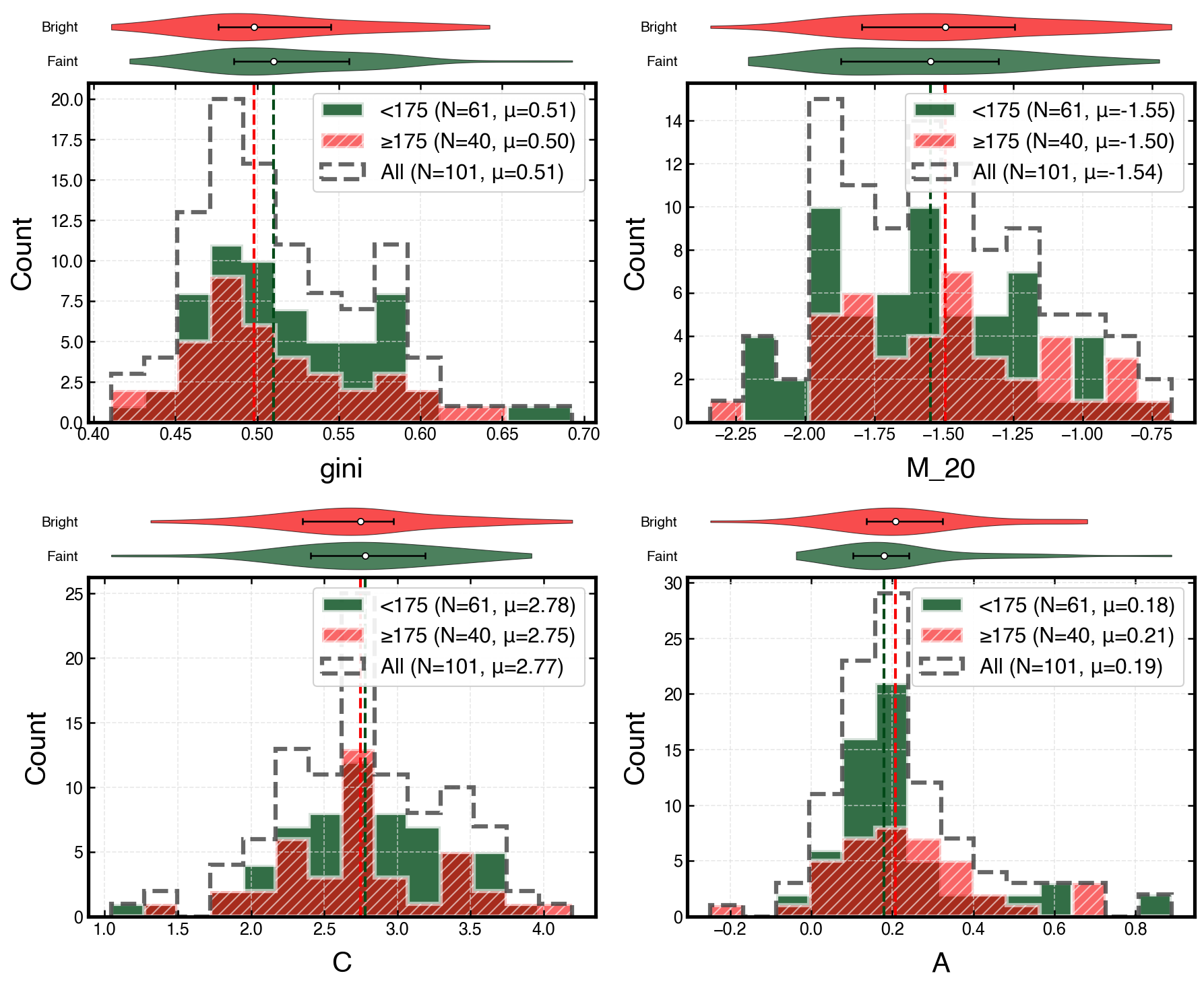}
    \caption{The non-parametric metrics of the bright(\texttt{SFR}$>175M_\odot~yr^{-1}$) and faint group(\texttt{SFR}$<175M_\odot~yr^{-1}$)}, marked by red and dark green respectively.
    \label{fig:non_parametric_metrics}
\end{figure}

\newpage
\section{Bar identification by residuals of single sersic fitting}

\begin{figure}[b]
    \centering
    \includegraphics[width=0.86\linewidth]{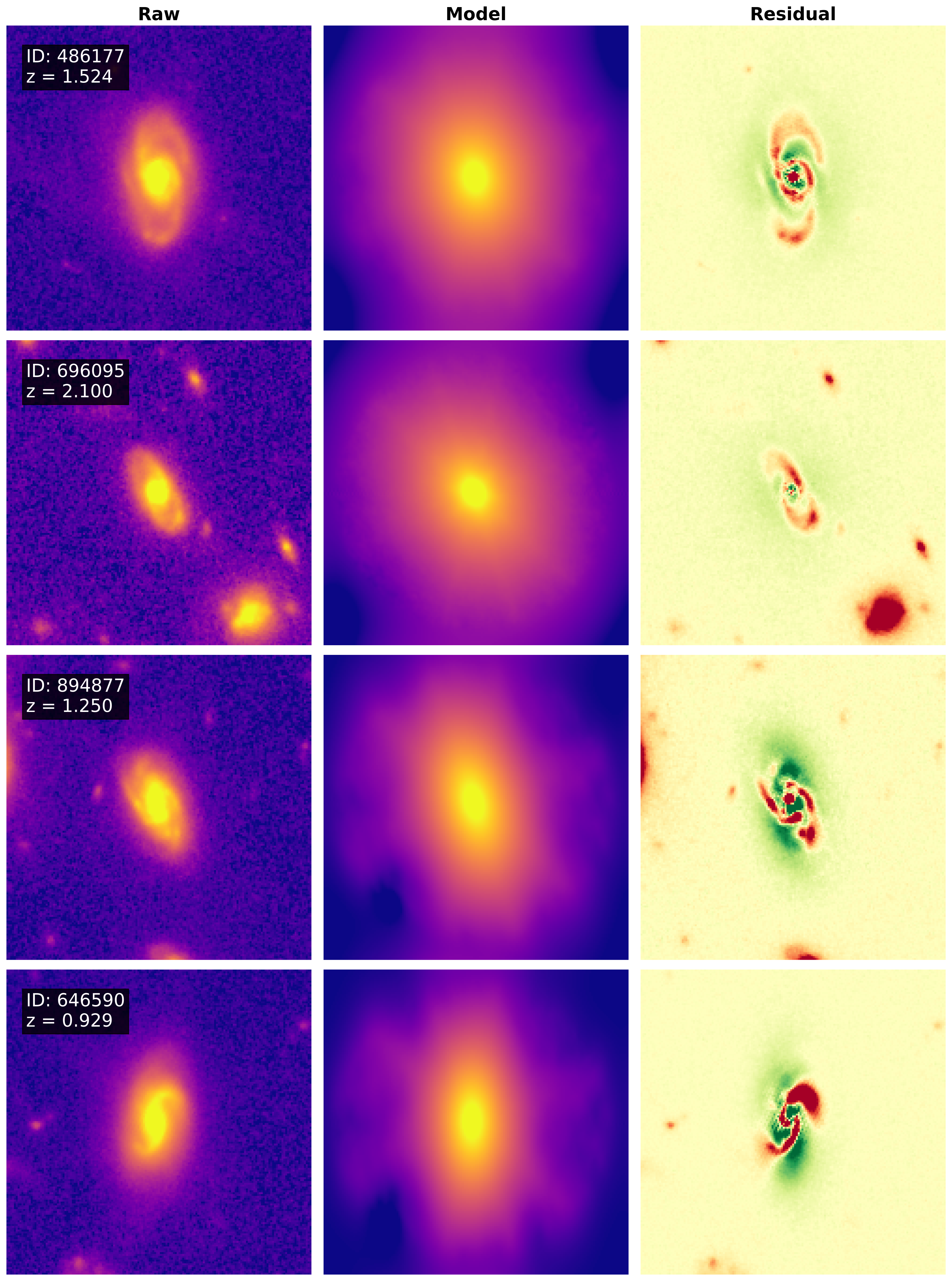}
    \caption{The confirmed bar candidates using the residuals of the single Sérsic fitting in band F277W. Each image is $6.5'' \times 6.5''$.}
    \label{fig:Bar}
\end{figure}

\begin{figure}[p]
    \ContinuedFloat
    \centering
    \includegraphics[width=0.86\linewidth]{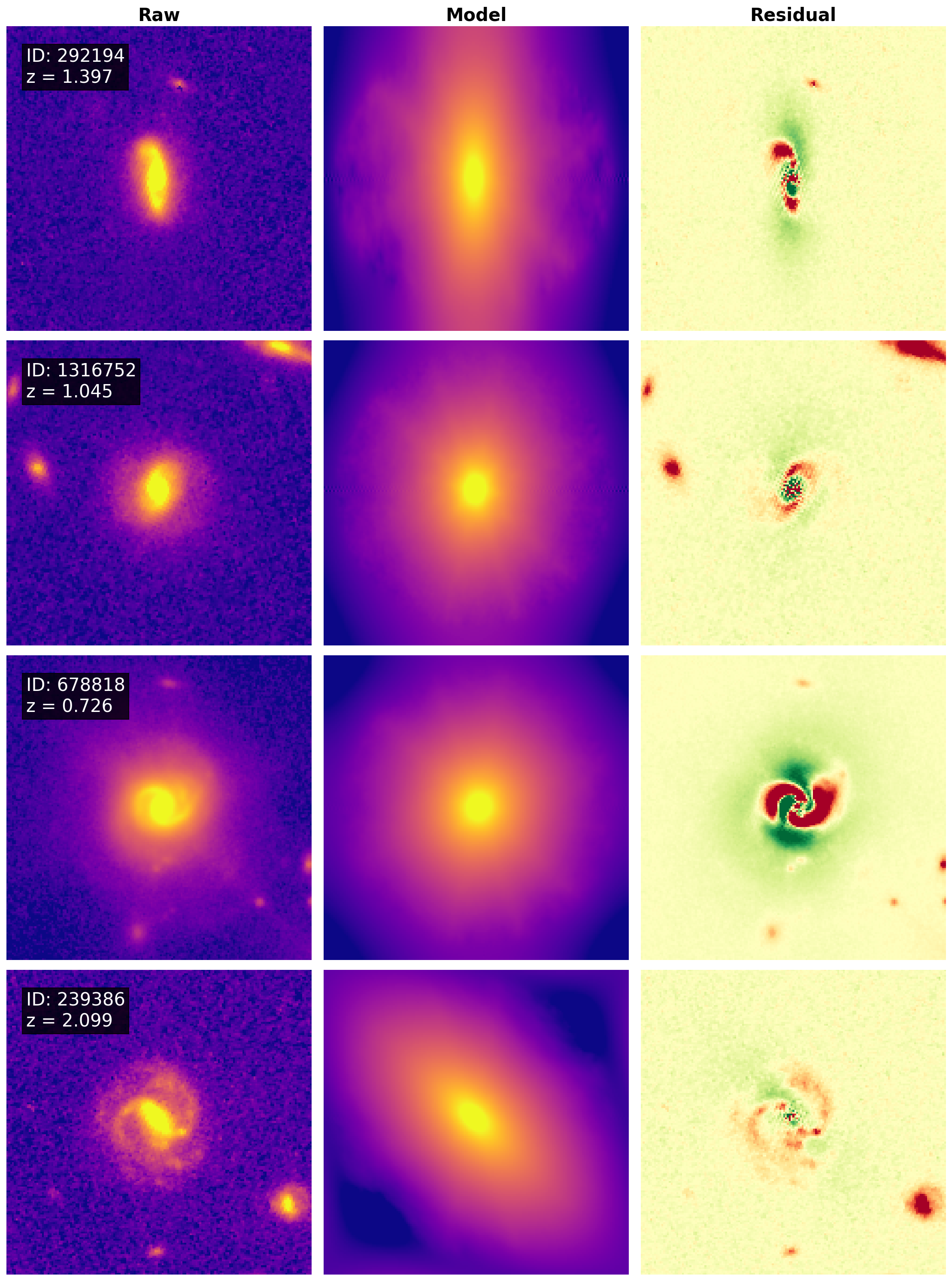}
    \caption{Continued.}
\end{figure}

\begin{figure}[p]
    \ContinuedFloat
    \centering
    \includegraphics[width=0.86\linewidth]{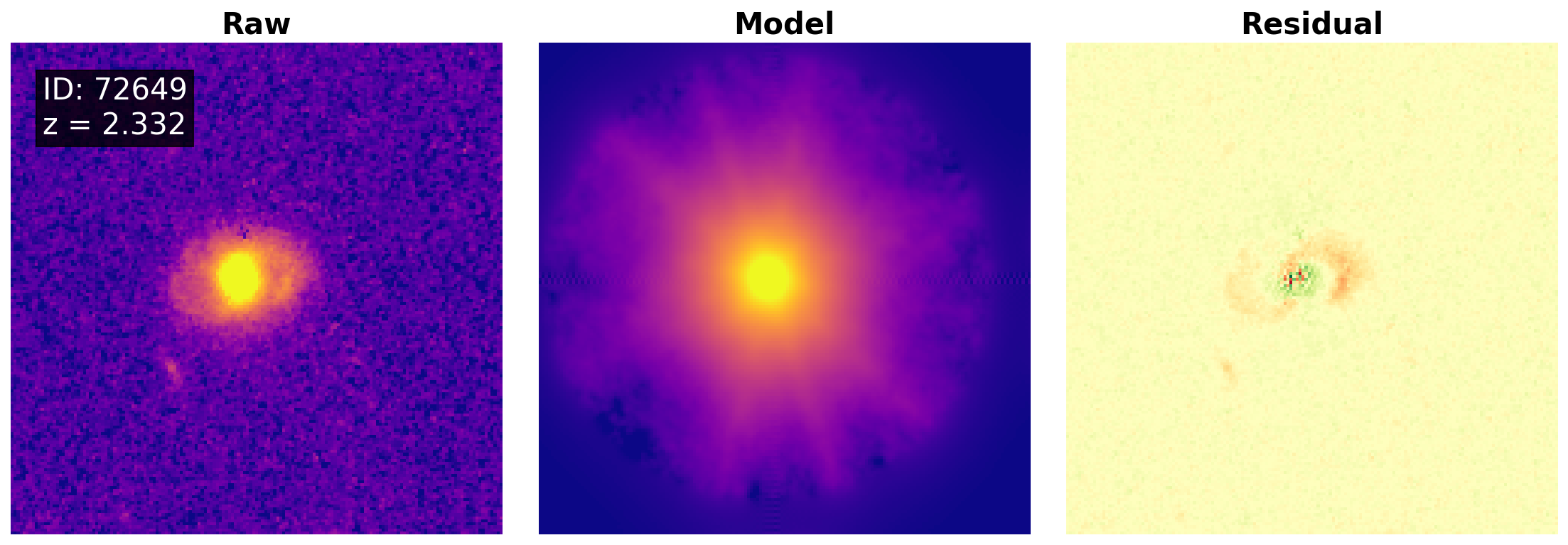}
    \caption{Continued.}
\end{figure}

\clearpage
\section{Physical properties and Coordinates}
\begin{center}
\small
\begin{longtable}{lccccccccc}
\caption{The catalog containing SMGs in the PRIMER COSMOS field with ALMA (A3COSMOS) physical properties and coordinates, alongside JWST (DJA) coordinates.} \label{tab:alma_jwst} \\
\toprule
\multicolumn{8}{c}{\textbf{ALMA (A3COSMOS)}} & \multicolumn{2}{c}{\textbf{JWST (DJA)}} \\
\cmidrule(lr){1-8} \cmidrule(lr){9-10}
ID & RA & Dec & $z$ & $\log_{10}(M_\ast/M_\odot)$ & $\log_{10}(L_\mathrm{dust}/L_\odot)$ & SFR & sSFR & RA & Dec \\
 & [deg] & [deg] &  & [dex] & [dex] & [$M_\odot$ yr$^{-1}$] & [Gyr$^{-1}$] & [deg] & [deg] \\
\midrule
\endfirsthead

\caption[]{(continued)} \\
\toprule
\multicolumn{8}{c}{\textbf{ALMA (A3COSMOS)}} & \multicolumn{2}{c}{\textbf{JWST (DJA)}} \\
\cmidrule(lr){1-8} \cmidrule(lr){9-10}
ID & RA & Dec & $z$ & $\log_{10}(M_\ast/M_\odot)$ & $\log_{10}(L_\mathrm{dust}/L_\odot)$ & SFR & sSFR & RA & Dec \\
 & [deg] & [deg] &  & [dex] & [dex] & [$M_\odot$ yr$^{-1}$] & [Gyr$^{-1}$] & [deg] & [deg] \\
\midrule
\endhead

\bottomrule
\endfoot
21931 & 150.11252 & 2.37655 & 1.92 & 11.77 & 12.22 & 165.96 & 0.28 & 150.11253 & 2.37654 \\
41140 & 150.18004 & 2.37828 & 4.91 & 10.45 & 12.65 & 446.68 & 15.85 & 150.18011 & 2.37831 \\
43882 & 150.07724 & 2.21141 & 2.18 & 11.06 & 11.72 & 52.48 & 0.46 & 150.07725 & 2.21141 \\
56056 & 150.10282 & 2.37939 & 1.85 & 11.49 & 11.96 & 91.20 & 0.30 & 150.10285 & 2.37940 \\
58907 & 150.07527 & 2.37943 & 2.00 & 10.88 & 11.86 & 72.44 & 0.96 & 150.07525 & 2.37941 \\
72649 & 150.18711 & 2.38011 & 2.33 & 11.38 & 12.06 & 114.82 & 0.48 & 150.18717 & 2.38014 \\
74827 & 150.13265 & 2.21185 & 2.11 & 11.62 & 13.10 & 1258.93 & 3.02 & 150.13266 & 2.21185 \\
75594 & 150.14361 & 2.47353 & 0.51 & 8.00 & 8.61 & 0.04 & 0.41 & 150.14361 & 2.47355 \\
77591 & 150.07574 & 2.38074 & 2.47 & 10.44 & 11.84 & 69.18 & 2.51 & 150.07584 & 2.38072 \\
77714 & 150.07715 & 2.38034 & 2.92 & 11.06 & 12.11 & 128.83 & 1.12 & 150.07717 & 2.38037 \\
82551 & 150.07594 & 2.21182 & 2.10 & 11.66 & 12.59 & 389.05 & 0.85 & 150.07595 & 2.21181 \\
83665 & 150.17324 & 2.46432 & 1.75 & 11.12 & 12.30 & 199.53 & 1.51 & 150.17325 & 2.46432 \\
86289 & 150.12178 & 2.21323 & 1.41 & 10.76 & 12.03 & 107.15 & 1.86 & 150.12179 & 2.21322 \\
97487 & 150.15865 & 2.46835 & 6.50 & 10.97 & 13.00 & 1000.00 & 10.72 & 150.15859 & 2.46835 \\
110346 & 150.16768 & 2.29873 & 2.09 & 11.54 & 12.26 & 181.97 & 0.53 & 150.16768 & 2.29875 \\
119950 & 150.10284 & 2.38455 & 1.95 & 11.23 & 12.02 & 104.71 & 0.62 & 150.10283 & 2.38451 \\
163763 & 150.03669 & 2.21790 & 1.82 & 11.71 & 12.36 & 229.09 & 0.45 & 150.03670 & 2.21788 \\
163962 & 150.18404 & 2.38637 & 2.13 & 11.20 & 11.95 & 89.13 & 0.56 & 150.18409 & 2.38638 \\
185349 & 150.12357 & 2.22070 & 2.82 & 10.12 & 11.87 & 74.13 & 5.62 & 150.12359 & 2.22070 \\
190486 & 150.07065 & 2.30516 & 2.94 & 10.45 & 12.07 & 117.49 & 4.17 & 150.07068 & 2.30512 \\
193219 & 150.05913 & 2.21988 & 1.14 & 11.03 & 12.03 & 107.15 & 1.00 & 150.05923 & 2.21980 \\
194663 & 150.15023 & 2.47515 & 1.45 & 12.00 & 13.16 & 1445.44 & 1.45 & 150.15023 & 2.47515 \\
201627 & 150.05481 & 2.22207 & 0.24 & 8.66 & 8.00 & 0.01 & 0.02 & 150.05464 & 2.22203 \\
219616 & 150.11419 & 2.22194 & 1.23 & 10.89 & 11.91 & 81.28 & 1.05 & 150.11436 & 2.22195 \\
239072 & 150.08426 & 2.21848 & 2.86 & 10.40 & 12.47 & 295.12 & 11.75 & 150.08432 & 2.21847 \\
239386 & 150.17771 & 2.14746 & 2.10 & 11.34 & 11.76 & 57.54 & 0.26 & 150.17773 & 2.14746 \\
239718 & 150.17688 & 2.14696 & 2.96 & 10.57 & 12.21 & 162.18 & 4.37 & 150.17694 & 2.14697 \\
255108 & 150.09940 & 2.40493 & 2.48 & 10.90 & 11.99 & 97.72 & 1.23 & 150.09941 & 2.40491 \\
264733 & 150.04226 & 2.22635 & 3.97 & 11.61 & 13.36 & 2290.87 & 5.62 & 150.04228 & 2.22636 \\
269152 & 150.13830 & 2.22500 & 1.78 & 11.50 & 11.73 & 53.70 & 0.17 & 150.13825 & 2.22500 \\
270203 & 150.13639 & 2.22533 & 1.48 & 10.75 & 12.32 & 208.93 & 3.72 & 150.13630 & 2.22524 \\
292194 & 150.02760 & 2.22796 & 1.40 & 11.44 & 12.27 & 186.21 & 0.68 & 150.02762 & 2.22797 \\
313094 & 150.10535 & 2.31290 & 2.28 & 11.45 & 12.72 & 524.81 & 1.86 & 150.10540 & 2.31284 \\
320337 & 150.08803 & 2.39510 & 3.42 & 10.51 & 12.49 & 309.03 & 9.55 & 150.08802 & 2.39510 \\
333216 & 150.11209 & 2.31401 & 2.11 & 11.24 & 11.90 & 79.43 & 0.46 & 150.11215 & 2.31400 \\
362264 & 150.16061 & 2.15627 & 2.07 & 10.70 & 12.00 & 100.00 & 2.00 & 150.16063 & 2.15627 \\
367935 & 150.14822 & 2.15696 & 1.52 & 9.29 & 10.91 & 8.13 & 4.15 & 150.14828 & 2.15692 \\
380083 & 150.07321 & 2.23324 & 2.33 & 10.56 & 12.15 & 141.25 & 3.89 & 150.07331 & 2.23325 \\
404147 & 150.09807 & 2.16577 & 1.15 & 11.56 & 12.29 & 194.98 & 0.54 & 150.09809 & 2.16577 \\
404453 & 150.09884 & 2.16627 & 1.13 & 11.26 & 12.06 & 114.82 & 0.63 & 150.09885 & 2.16623 \\
427795 & 150.12057 & 2.41809 & 5.64 & 11.82 & 12.48 & 302.00 & 0.46 & 150.12059 & 2.41809 \\
434557 & 150.18781 & 2.16064 & 1.86 & 11.08 & 12.46 & 288.40 & 2.40 & 150.18780 & 2.16065 \\
435921 & 150.09852 & 2.32084 & 2.70 & 10.37 & 12.21 & 162.18 & 6.92 & 150.09854 & 2.32087 \\
443595 & 150.11511 & 2.32127 & 2.13 & 10.68 & 12.13 & 134.90 & 2.82 & 150.11511 & 2.32127 \\
481073 & 150.16688 & 2.23582 & 1.84 & 10.94 & 11.96 & 91.20 & 1.05 & 150.16688 & 2.23582 \\
484109 & 150.17185 & 2.24071 & 2.09 & 10.85 & 12.56 & 363.08 & 5.13 & 150.17186 & 2.24070 \\
484771 & 150.14174 & 2.42569 & 2.82 & 11.01 & 12.32 & 208.93 & 2.04 & 150.14177 & 2.42568 \\
486177 & 150.18759 & 2.32248 & 1.52 & 11.73 & 12.14 & 138.04 & 0.26 & 150.18760 & 2.32249 \\
506425 & 150.10467 & 2.24366 & 1.91 & 11.33 & 12.09 & 123.03 & 0.58 & 150.10466 & 2.24367 \\
519264 & 150.10520 & 2.32581 & 1.26 & 9.82 & 11.64 & 43.65 & 6.58 & 150.10521 & 2.32582 \\
519984 & 150.13915 & 2.43197 & 3.07 & 11.63 & 12.58 & 380.19 & 0.89 & 150.13917 & 2.43198 \\
527216 & 150.11518 & 2.43330 & 0.02 & 6.00 & 8.00 & 0.01 & 10.00 & 150.11515 & 2.43333 \\
530894 & 150.09547 & 2.24667 & 2.18 & 10.82 & 12.33 & 213.80 & 3.24 & 150.09547 & 2.24668 \\
535014 & 150.13895 & 2.43377 & 2.60 & 11.31 & 12.97 & 933.25 & 4.57 & 150.13897 & 2.43376 \\
539347 & 150.12639 & 2.43285 & 1.27 & 11.11 & 12.16 & 144.54 & 1.12 & 150.12636 & 2.43287 \\
539550 & 150.15373 & 2.32797 & 4.37 & 11.58 & 13.16 & 1445.44 & 3.80 & 150.15375 & 2.32798 \\
572875 & 150.15365 & 2.24772 & 3.09 & 10.94 & 12.23 & 169.82 & 1.95 & 150.15365 & 2.24776 \\
580664 & 150.10634 & 2.25157 & 2.93 & 11.63 & 13.22 & 1659.59 & 3.89 & 150.10632 & 2.25157 \\
591583 & 150.12996 & 2.25271 & 2.93 & 11.28 & 12.64 & 436.52 & 2.29 & 150.12998 & 2.25268 \\
595185 & 150.10934 & 2.25277 & 2.01 & 9.97 & 11.56 & 36.31 & 3.92 & 150.10926 & 2.25268 \\
611293 & 150.10136 & 2.44270 & 2.24 & 10.45 & 11.73 & 53.70 & 1.91 & 150.10135 & 2.44270 \\
614294 & 150.13636 & 2.33352 & 1.08 & 10.93 & 11.27 & 18.62 & 0.22 & 150.13630 & 2.33362 \\
616447 & 150.15175 & 2.33459 & 1.74 & 11.84 & 12.10 & 125.89 & 0.18 & 150.15179 & 2.33460 \\
646590 & 150.09458 & 2.33633 & 0.93 & 11.17 & 11.78 & 60.26 & 0.41 & 150.09460 & 2.33635 \\
663586 & 150.13583 & 2.25792 & 4.45 & 9.63 & 11.64 & 43.65 & 10.30 & 150.13584 & 2.25789 \\
665972 & 150.06671 & 2.25773 & 2.10 & 10.87 & 11.35 & 22.39 & 0.30 & 150.06670 & 2.25773 \\
670959 & 150.07740 & 2.18536 & 2.20 & 11.10 & 12.10 & 125.89 & 1.00 & 150.07741 & 2.18536 \\
678818 & 150.14722 & 2.33717 & 0.73 & 11.10 & 11.68 & 47.86 & 0.38 & 150.14719 & 2.33718 \\
682011 & 150.09531 & 2.33895 & 3.13 & 10.61 & 11.92 & 83.18 & 2.04 & 150.09533 & 2.33898 \\
685092 & 150.11033 & 2.25844 & 2.45 & 9.82 & 11.29 & 19.50 & 2.94 & 150.11039 & 2.25864 \\
696095 & 150.10394 & 2.18642 & 2.10 & 11.68 & 12.69 & 489.78 & 1.02 & 150.10395 & 2.18642 \\
702416 & 150.07933 & 2.34058 & 2.68 & 10.66 & 12.55 & 354.81 & 7.76 & 150.07933 & 2.34056 \\
719895 & 150.07192 & 2.19020 & 2.80 & 11.51 & 12.40 & 251.19 & 0.78 & 150.07195 & 2.19018 \\
754141 & 150.10750 & 2.34435 & 2.69 & 10.48 & 11.48 & 30.20 & 1.00 & 150.10754 & 2.34434 \\
765691 & 150.08644 & 2.26432 & 1.19 & 9.48 & 12.04 & 109.65 & 36.14 & 150.08648 & 2.26430 \\
778981 & 150.12575 & 2.26662 & 0.72 & 8.65 & 11.13 & 13.49 & 30.06 & 150.12573 & 2.26659 \\
782304 & 150.10355 & 2.34608 & 1.03 & 11.01 & 12.00 & 100.00 & 0.98 & 150.10352 & 2.34608 \\
812701 & 150.03650 & 2.19837 & 3.00 & 10.99 & 11.99 & 97.72 & 1.00 & 150.03651 & 2.19838 \\
815643 & 150.03142 & 2.19637 & 3.50 & 10.66 & 13.17 & 1479.11 & 32.36 & 150.03142 & 2.19640 \\
817083 & 150.17488 & 2.35273 & 2.38 & 10.79 & 12.30 & 199.53 & 3.24 & 150.17487 & 2.35274 \\
819118 & 150.12487 & 2.26945 & 0.97 & 10.61 & 11.29 & 19.50 & 0.48 & 150.12489 & 2.26947 \\
854401 & 150.05385 & 2.20323 & 3.17 & 11.42 & 12.72 & 524.81 & 2.00 & 150.05385 & 2.20321 \\
865238 & 150.12286 & 2.36096 & 2.01 & 11.09 & 12.18 & 151.36 & 1.23 & 150.12290 & 2.36097 \\
877714 & 150.14127 & 2.27816 & 2.17 & 10.97 & 11.27 & 18.62 & 0.20 & 150.14129 & 2.27819 \\
885687 & 150.10877 & 2.20872 & 3.53 & 10.51 & 11.88 & 75.86 & 2.34 & 150.10876 & 2.20871 \\
889846 & 150.09596 & 2.36526 & 4.44 & 11.30 & 12.00 & 100.00 & 0.50 & 150.09597 & 2.36526 \\
892054 & 150.09855 & 2.36538 & 2.56 & 11.61 & 12.75 & 562.34 & 1.38 & 150.09856 & 2.36537 \\
893896 & 150.13087 & 2.20896 & 3.11 & 11.50 & 12.66 & 457.09 & 1.45 & 150.13088 & 2.20895 \\
894877 & 150.10925 & 2.20732 & 1.25 & 11.60 & 12.41 & 257.04 & 0.65 & 150.10925 & 2.20734 \\
911118 & 150.05975 & 2.28100 & 3.96 & 10.87 & 12.01 & 102.33 & 1.38 & 150.05975 & 2.28101 \\
923171 & 150.17076 & 2.36861 & 3.36 & 11.15 & 12.43 & 269.15 & 1.91 & 150.17080 & 2.36862 \\
931419 & 150.11433 & 2.36987 & 4.22 & 10.52 & 12.62 & 416.87 & 12.59 & 150.11432 & 2.36985 \\
941026 & 150.16353 & 2.37252 & 2.08 & 11.08 & 12.75 & 562.34 & 4.68 & 150.16353 & 2.37248 \\
944131 & 150.17157 & 2.28732 & 4.58 & 9.73 & 11.35 & 22.39 & 4.20 & 150.17158 & 2.28730 \\
949027 & 150.12448 & 2.28950 & 2.86 & 10.52 & 11.37 & 23.44 & 0.71 & 150.12448 & 2.28950 \\
954639 & 150.11234 & 2.37526 & 2.40 & 10.48 & 11.47 & 29.51 & 0.98 & 150.11236 & 2.37524 \\
956001 & 150.14415 & 2.37631 & 3.24 & 10.97 & 11.92 & 83.18 & 0.89 & 150.14418 & 2.37632 \\
1316680 & 150.10520 & 2.15661 & 1.58 & 10.91 & 12.14 & 138.04 & 1.70 & 150.10519 & 2.15664 \\
1316752 & 150.13932 & 2.14951 & 1.05 & 11.05 & 12.11 & 128.83 & 1.15 & 150.13932 & 2.14953 \\
1329539 & 150.05313 & 2.19807 & 1.33 & 11.28 & 12.22 & 165.96 & 0.87 & 150.05318 & 2.19808 \\
1334731 & 150.14477 & 2.21968 & 1.76 & 10.67 & 11.68 & 47.86 & 1.02 & 150.14482 & 2.21971 \\
1344523 & 150.09401 & 2.24580 & 1.77 & 10.62 & 12.21 & 162.18 & 3.89 & 150.09402 & 2.24586 \\
1351031 & 150.18430 & 2.26438 & 1.85 & 10.22 & 11.70 & 50.12 & 3.02 & 150.18428 & 2.26436 \\
1357333 & 150.19441 & 2.27327 & 0.37 & 8.72 & 10.50 & 3.16 & 6.07 & 150.19436 & 2.27331 \\
1357956 & 150.14728 & 2.28236 & 1.98 & 10.57 & 12.16 & 144.54 & 3.89 & 150.14729 & 2.28233 \\
1384974 & 150.14317 & 2.35598 & 1.52 & 10.22 & 12.17 & 147.91 & 8.91 & 150.14324 & 2.35602 \\
1393572 & 150.06144 & 2.37880 & 2.38 & 10.65 & 12.66 & 457.09 & 10.23 & 150.06146 & 2.37871 \\
1394965 & 150.06980 & 2.38154 & 2.30 & 9.24 & 11.46 & 28.84 & 16.52 & 150.06982 & 2.38152 \\
1435854 & 150.11169 & 2.48042 & 1.75 & 11.18 & 12.40 & 251.19 & 1.66 & 150.11175 & 2.48042 \\
1767929 & 150.07469 & 2.21648 & 1.92 & 11.31 & 12.56 & 363.08 & 1.78 & 150.07471 & 2.21648 \\
1768691 & 150.11126 & 2.40318 & 2.44 & 10.78 & 12.41 & 257.04 & 4.27 & 150.11130 & 2.40318 \\
1769008 & 150.10586 & 2.42881 & 2.99 & 11.21 & 12.84 & 691.83 & 4.27 & 150.10587 & 2.42881 \\
1769044 & 150.10697 & 2.42895 & 0.06 & 7.90 & 8.89 & 0.08 & 0.98 & 150.10698 & 2.42894 \\
1834187 & 150.10968 & 2.18797 & 2.30 & 11.37 & 12.48 & 302.00 & 1.29 & 150.10965 & 2.18794 \\
1834514 & 150.09824 & 2.16624 & 1.80 & 11.74 & 12.61 & 407.38 & 0.74 & 150.09825 & 2.16625 \\
1835014 & 150.12926 & 2.46431 & 3.77 & 11.63 & 13.03 & 1071.52 & 2.51 & 150.12929 & 2.46431 \\
1835211 & 150.14034 & 2.14715 & 4.20 & 11.70 & 13.10 & 1258.93 & 2.51 & 150.14033 & 2.14715 \\
1835575 & 150.17434 & 2.47825 & 0.54 & 9.17 & 8.39 & 0.03 & 0.02 & 150.17420 & 2.47810 \\
1835807 & 150.06508 & 2.26363 & 4.60 & 11.09 & 13.31 & 2041.74 & 16.60 & 150.06510 & 2.26363 \\
1836013 & 150.19625 & 2.17128 & 5.05 & 11.34 & 12.95 & 891.25 & 4.07 & 150.19626 & 2.17128 \\
1836571 & 150.17427 & 2.42973 & 6.35 & 11.20 & 13.17 & 1479.11 & 9.33 & 150.17429 & 2.42973 \\
1837147 & 150.10065 & 2.33485 & 5.90 & 11.04 & 13.02 & 1047.13 & 9.55 & 150.10064 & 2.33483 \\
1845716 & 150.10448 & 2.43538 & 6.20 & 11.35 & 13.31 & 2041.74 & 9.12 & 150.10446 & 2.43538 \\
1855826 & 150.09991 & 2.29711 & 4.00 & 11.39 & 13.22 & 1659.59 & 6.76 & 150.10009 & 2.29714 \\
1895614 & 150.10982 & 2.25775 & 5.85 & 11.35 & 12.97 & 933.25 & 4.17 & 150.10981 & 2.25779 \\
\end{longtable}
\end{center}


\clearpage
\bibliography{sample7}{}
\bibliographystyle{aasjournalv7}


\end{CJK}
\end{document}